\documentstyle[12pt,aaspp4]{article}

\slugcomment{To appear in the Astrophysical Journal}
\begin{document}
  
\def \cmm  {cm$^{-2}$ }
\def \kms {km~s$^{-1}$ }
\def \Lya {Ly$\alpha$ }
\def \Lyb {Ly$\beta$ }
\def \th {$\tau_{\rm{HeII}}$ }

\def \HI   {\ion{H}{1} }
\def \HeII {\ion{He}{2} }
\def \CIV  {\ion{C}{4} }

\def \object {HS 1946+7658 }

\title{Intrinsic Properties of the $<z> = 2.7$ Lyman Alpha Forest from
       Keck Spectra of QSO HS 1946+7658}
    
\author{DAVID KIRKMAN \altaffilmark{1} and DAVID TYTLER \altaffilmark{2}}

\affil{Department of Physics, and Center for Astrophysics and Space Sciences \\
	University of California, San Diego \\
	MS 0111, La Jolla, CA 92093-0111}

\altaffiltext{1}{dkirkman@ucsd.edu}
\altaffiltext{2}{dtytler@ucsd.edu}

\abstract 

We present the highest quality \Lya forest spectra published to date.
We have complete 7.9 \kms FWHM spectra between the \Lya and \Lyb
emission lines of the bright, high redshift (V=15.9, $z$=3.05) QSO HS
1946+7658. The mean redshift of observed \Lya forest clouds is $<z>=2.7$.
The spectrum has a signal to noise ratio per pixel of 2 \kms that
varies from 15 at 4190\AA \space to 100 at 4925\AA.  The absorption
lines in the spectrum have been fit with Voigt profiles, and the
distribution of Voigt parameters has been analyzed.  We show that
fitting Voigt profiles to high quality data does not give unique
results.  We have performed simulations to differentiate between true
features of the line distributions and artifacts of line blending and
the fitting process.  We show that the distribution of H~I column
densities is a power law of slope $-1.5$ from N(H~I) = $10^{14}$ \cmm
to N(H~I) $< 10^{12.1}$ \cmm.  We further show that our data is
consistent with the hypothesis that this power law extends to N(H~I) =
0, because lines weaker than N(H~I) $= 10^{12.1}$ \cmm do not have a
significant H~I optical depth.  At velocity dispersions between 20 and
60 \kms the velocity dispersion ($b$) distribution is well described
by a Gaussian with a mean of 23 \kms, and a $\sigma_b$ of 14 \kms.
Very similar N(H~I) and $b$ distributions were found at $<z> = 3.7$ by
Lu et al. (1997), indicating no strong redshift evolution in these
distributions for the \Lya forest.  However, our $b$ distribution has
a lower mean and a wider dispersion than in past studies at the same
redshift (eg. Hu et al. 1995) which had lower signal to noise spectra.
We unambiguously see narrow \Lya forest clouds with 14 \kms $\le b
\le$ 20 \kms that cannot be accounted for by noise effects.  Our data
also has absorption lines with $b \ge$ 80 \kms that can not be
explained by the blending of lower $b$ lines.  We find that the lower
cutoff in the $b$ distribution varies with N(H~I), from $b = 14$ \kms
at N(H~I) $= 10^{12.5}$ \cmm to $b = 22$ \kms at N(H~I) $= 10^{14.0}$
\cmm.  However, we see no similarly strong indication of a general
correlation between $b$ and N(H~I).  In contrast with previous
results, we find no indication of \Lya forest line clustering on any
scale above 50 \kms.  Even among lines with $10^{13.6} < N(H~I) <
10^{14.3}$ \cmm, which were previously thought to cluster very
strongly on velocity scales between 50 and 150 \kms, we see no
clustering on any scale above 50 \kms, although we do see a 
3 $\sigma$ clustering signal between 25 and 50 \kms among these higher
column density lines.  With the distributions we have
derived, we have calculated the expected He II optical depth of the
\Lya forest.  If there are no lines with N(H~I) $<
10^{12.1}$ \cmm, the \Lya forest is unlikely to provide a significant
portion of the He II optical depth at high redshift.  However, if the
distribution extends to N(H~I) $< 10^{9}$ \cmm, the \Lya forest can
provide all of the observed optical depth if N(He~II)/N(H~I) $\approx
100$.  We have calculated the redshift evolution of the optical depth
from the He~II \Lya forest based upon the line distributions we have
derived for the H~I \Lya forest.  If the \Lya forest is responsible
for the high redshift He~II optical depth and the spectral shape of
the UV background does not change with redshift, we predict \th
$\approx 2.4$ at $z=3.3$ to be consistent with the value of \th
previously found at $z=2.4$, provided that \Lya forest lines with
N(H~I) $<10^{13}$ \cmm evolve like those with N(H~I) $>10^{13}$ \cmm.

\section {Introduction}

The numerous \Lya absorption lines found in the spectra of high
redshift QSOs provide one of the few observational opportunities to
study the high redshift universe.  At low redshift, damped \Lya (DLA)
and metal line absorption systems are known to be
associated with galaxies (Lanzetta et al. 1995), and are believed to
be at high redshift as well.  The nature of the \Lya forest at high
redshift is unknown, though they are believed to have an intergalactic
origin.  Traditionally the absorption was thought to occur in tenuous
clouds confined either by pressure from a hot intergalactic medium
(IGM) (Sargent et al. 1980), or by gravity (Melott 1980).  However,
recent cosmological simulations indicate that the \Lya forest may be
composed of many different types of physical objects, including
filaments of warm gas, velocity caustics, and collisionally ionized,
high density, hot gas (Cen and Ostriker 1993, Hernquist et al. 1996).
A fluctuating \HI Gunn-Peterson effect may also cause broad absorption
features that would look like \Lya forest clouds (Miralda-Escude \&
Rees 1994).

\subsection {Column Density Distribution}

Many studies have attempted to determine the distribution functions
which characterize the \Lya forest.  The most recent (Hu et al. 1995)
found the column density distribution to be of the form
\begin{equation}
{{dn} \over {d\rm{N}}} \propto {\rm{N}}^{-1.46}
\end{equation}
from $\rm{N}=10^{12.3}$ \cmm to $\rm{N}=3 \times 10^{14}$ \cmm, where
N is the H~I column density in \cmm and $n$ is the number of clouds
per unit redshift.  Previous studies have found similar results, but
several found evidence for a break in the distribution in the vicinity
of $\rm{N}=10^{16}$ \cmm (Petitjean et al. 1993, and references
therein).  Line blending has prevented measurements of the
distribution below N(H~I) $= 10^{12.3}$ \cmm.

\subsection {Line Widths: Velocity Dispersion Distribution}

The form of the line width velocity dispersion ($b = \sqrt{2}
\sigma_v$) distribution has been discussed in a number of recent
papers. Pettini et al. (1990) reported that a majority of \Lya forest
clouds have velocity dispersions $b \le 22$ \kms and that $b$ is
correlated with N(H~I).  Carswell et al. (1991) and Rauch et
al. (1992) did not see a N(H~I)$-b$ correlation in Q1100-264 and
Q0014+813, respectively.  Rauch et al. (1993) reanalyzed the original
Pettini et al. spectrum of Q2206-199N and confirmed the appearance of
both the very narrow lines and the N(H~I)$-b$ correlation, but they
performed simulations that showed both were probably artifacts of low
signal to noise ratio (SNR).  Hu et al. (1995) examined the $b$
distribution of \Lya clouds in several objects and concluded that it
was described by a Gaussian with mean of 28 \kms, $\sigma_b =$ 10 \kms
and a cutoff below 20 \kms.  They agreed with the results of the
Hernquist et al. (1996) simulations that the broad absorption features
seen in spectra are blends composed of several narrower lines.  
Williger et al. (1994) and Lu et al. (1996) have also seen lines
with $b < 20$ \kms.

Low $b$ lines, if real, challenge the conventional view that the \Lya
forest clouds are hot and highly ionized.  It is possible to obtain
the low temperatures implied by these $b$ values -- $b = 20$\kms is
equivalent to 24,000 K -- if the clouds are very dense, $n_{H~I}
\approx 0.1 \rm{cm}^{-3}$ (Pettini et al. 1990).  However, the
inferred size of absorbers with this density ($\approx 1$ pc) is
difficult to reconcile with the large sizes ($\approx 100 \rm{h}^{-1}$
kpc) inferred from absorption in the spectra of close QSO pairs
(Smette et al. 1991, Dinshaw et al. 1995).

Other physical mechanisms which could produce low temperature \Lya forest
clouds have been proposed.  \Lya forest clouds can be cooled to low
temperatures and still have a large size if they are adiabatically
expanding against a confining force, perhaps gravity from dark matter
halos or pressure from a hot IGM (Duncan et al. 1991). Further work on
this model (Petitjean et al. 1993) has shown that the clouds are
composed of shells of gas, with the high temperature gas toward the
center of the cloud.  For random lines of sight through the cloud,
there is a 20 \% chance of passing through a thin outer layer of the
cloud with T $< 15,000$K, which corresponds to a $b$ of at least 15.7
\kms; where the observed $b$ for such a cloud will be higher if there are
turbulent motions.  Giallongo and Petitjean (1994) have pointed out
that clouds as large as 20 kpc can be cooled to about 20,000K ($b$ =
18.1 \kms) by inverse Compton scattering of electrons on the cosmic
microwave background.  In the N-body cold dark matter simulations
performed by Hernquist et al. (1996), most of the \Lya forest
absorption occurs in structures with low temperatures ($T \approx
20,000 $K), and the apparent width of the lines is due entirely to the
blending of many components.

\subsection{Significance of the \Lya forest}

Knowledge of the distributions which describe the \Lya forest cloud
population is critical not only to understand the nature of the clouds
themselves, but also to calibrate other astronomical observations.
[D/H] measurements depend upon the line distributions to determine the
likelihood that an H~I line will masquerade as Deuterium in a given
system (Tytler and Burles 1996).  Understanding the column density
distribution at low columns ($N < 10^{12.5}$) is crucial in the
interpretation of the recent measurements of the He~II optical depth at
high redshift (Jakobsen et al. 1994, Davidsen et al. 1996).  If there
are a large number of absorbers with $N < 10^{12.5}$ \cmm, \Lya forest
clouds can account for all of the observed He~II absorption for
expected values of [He~II/H~I], without any need for additional
Gunn-Peterson absorption.  In a broader sense, it now appears that
most baryons are in the structures which show \Lya forest absorption
(Meiksin and Madau 1993, Rauch et al. 1997, Weinberg et al. 1997).
Comparison of the observed forest with the results of computer
simulations constrain the cosmological baryon density, $\Omega_b$, and
also the thermal and ionization history of the universe on the largest
scales.
 
All wavelengths presented in this paper are vacuum values, and all
observed wavelengths have been reduced to the solar rest frame.  We
use $H_0 = 100\rm{h}$ \kms Mpc$^{-1}$.

\section {Observations and Extraction}

In July 1994, we used the HIRES spectrograph (Vogt 1994) on the Keck
Telescope to take seven spectra of QSO HS 1946+7658 (V=15.9, Z=3.051)
discovered in the Hamburg Survey (Hagen et al. 1992).  We selected
this object because it is the brightest high redshift QSO known.  All
seven observations were made during good seeing conditions, with a
1.148" slit, the C5 decker, and 1x2 on-chip binning (along the slit)
on a Tex 2048 CCD.  The total integrated time on the object was 13.3
hours.  We have complete wavelength coverage between the \Lya and \Lyb
emission lines.  The spectra were optimally extracted and wavelength
calibrated using the EE package written by Tom Barlow.  Once
extracted, the individual spectra were rebinned onto a common linear
wavelength scale, and combined using weights to form a composite
spectrum.  The resulting spectrum has a SNR per pixel ($\approx 0.031$
\AA) that varies from 15 at 4190\AA \space to 100 at 4925\AA.  The SNR
also varies across each echelle order, being a factor of $\approx 1.5$
greater in the center of the order than at the edges.  The resolution
of the spectrum is 7.9 \kms.  The spectrum was not fluxed to an
absolute energy scale.  A preliminary local continuum level was
obtained independently for each echelle order by identifying regions
in the spectrum unaffected by absorption features and interpolating
between them with a seventh order Legendre polynomial, which provide
good fits to the continuum level in the echelle orders redward of \Lya
emission, where there is little absorption.  The spectrum between the
\Lya and \Lyb emission lines is displayed in Figure 1.

A DLA line that fell between two orders of the echelle image prevented
us from obtaining an accurate continuum level and line profiles in the
region between 4664 \space \AA \space and 4721 \space \AA.  This
region of the spectrum is not included in Figure 1, and none of the
lines in that region have been included in any of our analysis.  We do
not discuss the DLA, which has been extensively studied by others (Fan
\& Tytler 1994, Lu et al. 1995).  Two regions of the spectrum
presented in Figure 1 have not been fit with Voigt profiles.  The
region between 4635 \space \AA \space and 4640 \space \AA \space is
believed to contain both \Lya forest lines and Si III (1206.5) from
the $z=2.84$ DLA system.  Because it is heavily blended and we have no
other Si III transition available to guide us in fitting, we have
excluded the region from our analysis.  No data was obtained near
4300.5 \space \AA \space because of the ink spot on the HIRES CCD.

\subsection {Fitting Line Parameters}

We used Carswell's VPFIT software to fit all of the absorption
features with Voigt profiles.  The Voigt parameters and
identifications for each absorber in the spectrum are listed in Table
1.  We broke the spectrum up into the smallest possible absorption
regions, bounded where the spectrum reached the continuum level.
Initial line parameters (guesses) for each region were set by hand and
given to VPFIT, which adjusts the initial parameters to minimize the
$\chi^2$ between the fit and the data.  We added lines until we
obtained a reduced $\chi^2$ between fit and data of less than 1.3 over
the region being fit, and we always attempted to use the fewest number
of lines capable of meeting this condition.  P($\chi^2$), the
probability of good fit being rejected because of extreme noise
fluctuations, is between 1\% and 10\% for $\chi^2 < 1.3$, depending
upon the number of lines and the size of the region being fit.

Because the local continuum level is not well determined in the \Lya
forest, we adjusted the continuum level in each order to give an
optimum fit between the data and the calculated Voigt profiles.  The
continuum level of each order was taken to be a Legendre polynomial --
adjustments of the continuum level were made by changing the
coefficients of the polynomial.  The polynomial describing each order
was constrained to be of order seven or less.

The Voigt profiles obtained by this fitting procedure may not
accurately describe the absorbing gas.  For example, several discreet
clouds located tens of kpc apart, each with a small $b$ value, may
blend together in the spectrum and appear in the line list as a single
localized cloud with a high $b$ value. This ambiguity complicates the
process of determining the physical properties of the absorbing gas.
However, the VPFIT procedure usually results in a consistent
description of the spectra and is universally used (Carswell et al.,
1984).  Unfortunately, it does not always yield unique results.
(Section 3.2)

\subsection {Metal Lines}

We have attempted to identify all of the metal lines present in the
\Lya forest.  We began by searching for C IV (1548.195, 1550.770)
doublets, both in and out of the forest.  We also searched for any
line associated with known absorption systems identified redward of
the forest (e.g. the DLA systems at $z=2.844$ and $z=1.738$), as well
as for the Si~IV (1393.755, 1402.770) and Mg II (2796.352, 2803.531)
doublets.  This method identified all of the metal line systems
previously in the literature (Fan and Tytler 1994, Lu et al. 1995) as
well as several new systems (C IV systems at $z_{\rm{abs}}=$ 2.228,
2.465, 2.499, 2.985, and 3.027).  All absorption lines short of \Lya
emission at 4924 \AA \space were assumed to be \Lya unless we found
other identifications.  Because many of the lines in the forest are
blended, line wavelengths have a much greater uncertainty than they do
outside of the forest and we may have misidentified a few metal lines
as \Lya.

\section {Absorption Line Parameters}
\subsection {Line Blanketing and Blending} 

We must distinguish between an observed distribution obtained by
fitting Voigt profiles to absorption features in a spectrum, and an
intrinsic distribution that describes the physical state of the gas
responsible for the absorption.  The two are different because of line
blanketing and line blending.

Line blanketing occurs when a weak line is located in the absorption
profile of a stronger line, and as a result is not observed.  The
stronger line does not have to be saturated to effectively blanket
weaker lines.  Line blanketing decreases the observed number of low column
density lines.

Line blending occurs when two lines can not be resolved
and are instead fit with a single Voigt profile.  The
effect of line blending is to create a higher proportion of wide lines
than is present in the intrinsic distribution.  Both effects decrease
with increasing SNR.

\subsection {Non-uniqness of fitted Line Parameters}

The process
of fitting Voigt profiles to absorption spectra can not yield unique
results.  Examples of this are shown in Figure 2, where we show two
different sets of Voigt profiles which fit each absorption feature.
In both cases, although one of the solutions is better than the other,
both meet our criterion for a satisfactory fit.  The reduced $\chi^2$
for the region shown in each panel is as follows: Panel A -- 0.83,
Panel B -- 1.15, Panel C -- 1.04, Panel D -- 1.16.  The differences
between line parameters in the different solutions are far greater
than the formal fitting errors for those parameters.  We do not have
the resources to systematically search all of parameter space during
the fitting process, and in general we accept the first satisfactory
solution found.  As can be seen in Figure 2, this solution can be
distant in parameter space from better (lower $\chi^2$) solutions.
  
We do not believe that the fitting process can be made more unique by
lowering the $\chi^2$ acceptance threshold.  $\rm{P}(\chi^2)$, the
probability that random noise will cause the fit to be rejected is
between 1\% and 10\% (depending upon the size of the region being
fitted and the number of lines in the fit) for our $\chi^2$
threshold of 1.3.  If noise causes the fit to be rejected, more lines,
usually heavily blended, will be added.  If we attempt to combat the
non-uniqueness of the fitting process by lowering the acceptable
$\chi^2$ for a fit, we will systematically increase the number of
lines observed in the spectrum, and because most of the additional
lines will be heavily blended, they are not likely to be well
determined, and the new solution will be less unique.

The observed distribution functions for the H~I \Lya forest towards
QSO HS 1946+7658 can be obtained directly from Table 1, and are shown
in Figures 3, 4, 5, and 6.  The intrinsic distributions are more
interesting because they must be used when comparing different
observations or when comparing observation with theory.  The observed
distributions obtained from different observations should not be
directly compared with each other because they depend upon both the
SNR of the data and the details of the process used to fit Voigt
profiles to the data, among other things.

Unfortunately, it is not possible to invert the problem and compute
the intrinsic line distribution responsible for the absorption
spectrum from an observed distribution and the details of the
observation. With the aid of simulations, however, it is possible to
determine intrinsic distributions from observed distributions.

Throughout the rest of the paper we will be talking about three types
of distribution.  An intrinsic distribution describes the physical
state of the absorbing gas -- either in reality or as assumed for a
simulation.  An observed distribution is derived from lines fit to
real QSO absorption line spectra.  Simulated-observations give
distributions that have been derived from simulated spectra, which we
treated like the real spectra.

\subsection {Simulations and the Intrinsic Line Distributions}

We have performed simulations to determine the intrinsic distributions
of the \Lya forest lines in our spectrum.  A simulated-observation
consists of creating an artificial spectrum from a given intrinsic
distribution, and then fitting the artificial data in the same manner
as the real data.  The intrinsic N(H~I) and $b$ distributions we used
were the distributions Hu et al. (1995) found to describe the \Lya
forest.  (See sections 3.4 and 3.5 for details).  We attempted to
mimic every systematic aspect of the real spectra.  The artificial
spectrum was convolved with a Gaussian to simulate the 7.9 \kms
resolution of the real data.  It was broken up into echelle orders at
the same wavelength boundaries as the real data.  Of the 13 echelle orders
observed towards \object, 10 were simulated.  For each order in
the real spectrum, we fit the noise array with a 4'th order Legendre
polynomial and then added noise to the simulated data at the level of
the smoothed noise array.  Voigt profiles were fit to the simulated
spectra with the same procedure used to fit the real spectra.  The
simulated spectra was not continuum fit as the real data, because the
continuum errors in the real spectrum are not known.  Since continuum
errors are likely to show up as lines with very high velocity
dispersions ($b >$ 100 \kms), the simulated-observations may not yield
the correct intrinsic distribution of such lines.

\subsection {N(H~I) Distribution}

We find the intrinsic N(H~I) distribution per unit redshift
to be
\begin{equation}
{{dn} \over {d\rm{N}}} = \cases{
	0  & N $<$   N$_{min}$ \cr
	6.2 \times 10^8 \rm{N}^{-1.5} & N $\ge$ N$_{min}$ } 
\end{equation}
with N$_{min} < 10^{12.1}$ \cmm.  The N(H~I) distribution observed
towards \object is shown in Figure 3.  All observed \Lya lines
(including those associated with metal line systems) were used to
determine this distribution.  The mean redshift of these observations
is 2.7.

The N(H~I) distribution from the simulated-observations is shown in
Figure 7.  The intrinsic distribution used to make the simulation is
given by Equation 2, with N$_{min} = 10^{12.0}$ \cmm.  This is similar
to the N(H~I) distribution used by Hu et al. (1995), but we have
adjusted the normalization to give the number of lines that we
observed at N(H~I) $< 10^{13}$ \cmm.  Figures 3 and 7 are the same to
within error bars, so we conclude that intrinsic distribution of the
\Lya forest clouds is given by Equation 2.  The simulated-observations
indicate that the entire fall off seen in Figure 3 is expected from
line blanketing, and that our line list is complete for
N(H~I)$>10^{13}$ \cmm.

Because the low end of the N(H~I) distribution can have a significant
or even dominant effect upon the mean He~II optical depth of the \Lya
forest clouds, we wish to determine the maximum N$_{min}$ that is
consistent with our observations.  We performed a number of different
partial simulations, corresponding to orders 74 and 73 in the spectrum
of HS 1946+7658, with N(H~I) distributions characterized by different
N$_{min}$.  Orders 74 and 73 cover the wavelength region from 4783 \AA
\space to 4919 \AA.  We only performed partial simulations so that we
could perform many in a reasonable time, and we simulated orders 74
and 73 because they had the highest SNR and hence the most low N(H~I)
lines.  The results of these simulations indicate that any N$_{min}$
below $10^{12.1}$ \cmm is acceptable.  Any value of N$_{min}$ above
$10^{12.1}$ \cmm results in simulated-observations that contain too
few lines with N(H~I) $<10^{12.1}$ \cmm.  If N$_{min}$ is set below
$10^{12.1}$ \cmm, the H~I optical depth of the additional lines is so low
that they have no effect upon the observed distribution for data with
our SNR levels.  In Figure 8 we show both the mean H~I absorption of
the \Lya forest at a redshift of 2.7 for different values of
N$_{min}$, as well as the mean differential absorption of the H~I \Lya
forest per unit column density. Note that most of the H~I \Lya forest
opacity occurs in clouds with $10^{13} < \rm{N(H~I)} < 10^{15}$ \cmm,
and that the differential contribution to the \Lya opacity is a
maximum at N(H~I) $=10^{14}$ \cmm.  Also note that the contribution to
$\rm{D_A}$, the mean drop below the continuum level due to \Lya
absorption, from lines below N(H~I) = $10^{12.1}$ \cmm would be less
than 2 \%.

It is physically reasonable for the N(H~I) distribution to extend to
N(H~I) $= 10^{12.1}$ \cmm.  If the distribution of column densities is
given by Equation 2, the mean proper distance between a \Lya forest
cloud and it's nearest neighbor with N(H~I) $\ge N_0$ is given by (in
a $\Omega = 1$, $q_0 = 0.5$ cosmology)
\begin{equation}
d = {100 h_{100}^{-1} {\rm{N}_{min} \overwithdelims () 10^{12.07}}^{1 \over 2} \rm{kpc}}.
\end{equation}
If the size of the all of the \Lya forest clouds is 100h$^{-1}$ kpc, the
filling factor of the \Lya forest approaches unity when N$_{min} \approx 
10^{12.1}$ \cmm.

\subsection {$b$ Value Distribution}

We made simulated spectra from the intrinsic distribution that Hu et
al. (1995) found to describe the \Lya forest: a Gaussian with a mean
of 28 \kms and $\sigma_b = 10$ \kms.  The $b$ distribution from the
simulated-observations for lines with N(H~I) $>10^{12.5}$ \cmm is
shown in Figure 9.  The simulated-observations clearly resemble the
intrinsic distribution, but there are significant differences between
the two.  Primarily, there is a tail of wide $b$ lines in the
simulated-observations that were not present in the intrinsic
distribution -- these are the lines in the ``hump'' seen in the
N(H~I)-$b$ distribution between $10^{12.5}<$N(H~I) $<10^{13.5}$ \cmm
(See section 3.6).  The simulated-observations are not a Gaussian, but
if we force a Gaussian fit upon the region between $b=20$ and $b=60$
\kms we find a mean of 29 \kms and a $\sigma_b$ of 9 \kms, which is
nearly identical to the intrinsic distribution used to make the
simulation.

The simulated-observations look a lot like the real observations,
except that the position of the peak of the $b$ distribution occurs at
a significantly lower $b$ value in the real observations.  This means
that the intrinsic $b$ distribution used to create the
simulated-observations is not the intrinsic $b$ distribution of the
real data.  We wish to determine the intrinsic $b$ distribution of the
real data, but the only true way to do this is guess the correct
intrinsic distribution, perform simulations, and then see that the
simulated-observations are the same as the real observations.
Unfortunately, simulations are extremely expensive to perform.

To avoid performing more simulations, we have simply noted that if we
fit the $b$ distribution from the simulated-observations between 20
and 60 \kms with a Gaussian, we get a result that closely resembles
the intrinsic distribution used to create the simulation.  In short,
the simulated-observations indicate that the observed $b$ distribution
will closely resemble the intrinsic distribution.  However, due to
line blending, the observed distribution will have a tail of high $b$
lines that is not present in the intrinsic distribution.  The simulations
performed by Hu et al. (1995) and Lu et al. (1997) displayed the
same behavior.

If we do the same thing to our observed $b$ distribution -- that is
fit it with a Gaussian between $b=20$ and $b=60$ \kms, we get a mean
for the Gaussian of 23 \kms and a $\sigma_b$ of 14 \kms.  As was the
case for the simulated-observations, this Gaussian does not fit the
observed distribution perfectly .  However, based upon the results of
the simulated-observations we believe that this is close to the actual
intrinsic $b$ distribution for the \Lya forest.  Without further
simulations we will not be able to determine the exact form of the
intrinsic $b$ distribution, but for the purposes of this paper it is
sufficient to know that it is close to a Gaussian with a mean of 23
\kms and a $\sigma_b$ of 14\kms.  The mean $b$ is lower and the
$\sigma_b$ is larger than found by Hu et al. (1995) with lower SNR
data.  We do not know whether the difference comes from the data,
simulations, or the fitting.

There is a significant non-Gaussian tail to the $b$ distributions from
both the real data and the simulated-observations.  In the case of the
simulated-observations, it is due entirely to line blending.
Consequently, line blending must create a tail in the observed
distribution as as well, although the observed tail may contain more
lines than we would expect pure line blending.  3.4 \% (16 of 466) of
the lines observed towards \object have $b > 80$ \kms, whereas only
0.5 \% (2 of 363) of the lines in the simulated-observations have $b >
80$ \kms.  The extra lines may be indicative of a non-Gaussian tail
that is present in the intrinsic $b$ distribution.  Unfortunately, we
would have to make new simulations with a mean of 23 \kms and a
$\sigma_b$ of 14 \kms to verify this.  Some of the high $b$ lines
could be removed from the spectrum with a high order continuum
(Legendre polynomial of order $\ge$ 15), but we chose not to do this
because the continuum levels of the 16 red orders in this spectrum
which have no \Lya absorption are all well fit by seventh order
Legendre polynomials.

Physically, an intrinsic population of high $b$ \Lya lines could
represent high temperature gas ($10^6$ K) such as is found in the
halos of galaxies.  If these lines are hot \Lya clouds, they may show
high ionization metals such as O VI, Ne VIII, Mg X or Si XII (Verner,
Tytler, \& Barthel 1994). At such high temperatures we do not expect
to see strong C IV absorption, which reaches its maximum abundance at
T=$1.1 \times 10^5$ K, and begins to decrease rapidly at higher
temperatures.  We do not have high enough SNR at blue wavelengths to
see O VI in this spectrum.

\subsection {N(H~I) - $b$ Distribution and the lower cutoff in the $b$
	     distribution}

We have found a correlation between the lowest $b$ values and N(H~I).
In Figure 5 we have plotted N(H~I) vs. $b$ for the \Lya lines observed
towards \object.  The correlation between the lowest $b$ values,
denoted $b_{min}$, and N(H~I) is easily seen among the lines with
N(H~I) $> 10^{12.5}$ \cmm, and is approximately given by the line
\begin{equation}
b_{min} = 14 + 4 \times \rm{Log} { \rm{N(H~I)} \overwithdelims () 10^{12.5}}
	\, \rm{km} \, \rm{s}^{-1}, 
\end{equation} 
which is shown in Figure 5.  This correlation between $b_{min}$ and
N(H~I) was not present in the simulated-observations, which are
plotted in N(H~I) vs.  $b$ format in Figure 10.  The intrinsic
distribution used in the simulated-observations had a lower cutoff of
20 \kms, which was perfectly preserved in the simulated-observations
for lines with N(H~I) $> 10^{12.5}$ \cmm.  Because the lower cutoff in
the simulated-observations was so well preserved for these lines we
believe that Equation 4 represents a correlation present in the real
intrinsic distribution.

Note that the N(H~I)-$b$ distribution from the simulated-observations
is very similar to the observed N(H~I)-$b$ distribution.  Both show a
``hump'' containing a number of wide $b$ lines between N(H~I) =
$10^{12.5}$ and $10^{13.5}$ \cmm, and a number of low $b$ lines with
N(H~I) $<10^{12.5}$ \cmm.  Also note that the intrinsic distribution
used to create the dataset from which the simulated observations were
made (Figure 11) is essentially featureless.

There are three conspicuous lines in Figure 5 which lie below $b$ = 20
\kms and above N(H~I) = $10^{12.5}$ \cmm.  We suspect that these lines
may be unidentified metals, but we can not positively identify them
because \Lya forest absorption may be blanketing other transitions.
Possible non-\Lya identifications for the lines are as follows.  The
line at 4190.2 \AA \space is definitely not part of any of the C IV
(1548.195, 1550.77), SiIV (1393.755, 1402.77) or N V (1238.821,
1242.804) doublets, it could be either Mg II (2796.352) or Al III
(1854.716).  The line at 4176.9 \AA \space is not part of the C IV
(1548.195, 1550.77), or N V (1238.821, 1242.804) doublets, it could be
Mg II (2796.352), Mg II (2803.531), Al III (1854.716) or Al III
(1862.79). The line at 4293 \AA \space could be C IV (1548.195).
However, all three lines appear as \Lya in our line list.

We see twice as many lines with $b<20$ \kms and N(H~I) $< 10^{12.5}$
\cmm in the observed distribution as we did in the
simulated-observations, indicating that any cutoff present in the $b$
distribution for lines with N(H~I) $< 10^{12.5}$ is below 20 \kms.
This is what we expect if Equation 4 is extrapolated below
N(H~I) $= 10^{12.5}$ \cmm.  This region contains 10 \% of the observed
lines, compared to 4.7 \% in the simulated-observations, which used a
lower cutoff in the intrinsic $b$ distribution of 20 \kms.  The 4.7 \%
must be noise features because there were no lines with $b<20$ put
into the simulation.  The formal fitting errors are not large enough
(typically $\approx 20$\% in $b$) to explain the excess.  The
remaining 5.3 \% are apparently real.  Rauch et al. (1992) first
realized while reanalyzing the spectrum of Q2206-199N that the
observed distribution will contain low $b$ lines at low N(H~I)
(relative to the quality of the data) that are not present in
intrinsic distribution.

While the correlation we see between $b_{min}$ and N(H~I) is
reminiscent of the correlation between $b$ and N(H~I) found by Pettini
et al. (1990), it is not the same.  The correlation reported by
Pettini et al. (1990) was a general correlation between every observed
$b$ and every observed N(H~I), a correlation that is easily seen in
the observed distribution of \object, and in the distribution from the
simulated-observations.  The correlation in the low SNR data was later
shown by Rauch et al. (1993) to be an artifact of the fitting process,
a conclusion that we verify.  The origin of this correlation can be
seen from the simulated-observations, where it occurs for two reasons:
there are a large number of high $b$ lines with
$10^{12.5}<\rm{N(H~I)}<10^{13.5}$ \cmm (due to line blending), and a
large number of low $b$ lines with $\rm{N(H~I)} < 10^{12.5}$ \cmm.  We
do not understand the mechanism that produces low $b$ lines below
$10^{12.5}$ \cmm.  They are not all pure noise features because their
SNR is too high (a N(H~I) $=10^{12}$ \cmm line has an equivalent width
of 20 mA at $z=2.7$, which is about a 5 $\sigma$ event for our data.)
However, because they appear in the simulated-observations we know
that the observed distribution contains lines like this that are not
part of the intrinsic distribution.

While we see no general intrinsic correlation between N(H~I) and $b$,
we are reporting a correlation between the very lowest $b$ values and
N(H~I) that is present in the observed distribution and in the
intrinsic distribution. And this correlation, along with the fact that
the very narrowest lines have $b$ values between 14 and 20 \kms,
raises the same issues originally brought up by Pettini et al (1990).
How is it that lines can obtain $b$ values of $< 20$ \kms, and why
should there be a correlation between the very smallest $b$ values and
N(H~I)?

There is (at least) one mechanism that can cool a \Lya forest cloud to
14 \kms.  In the adiabatically expanding \Lya cloud model discussed by
Duncan, Vishniac \& Ostriker (1991) and refined by Petitjean et al.
(1993), the \Lya clouds are adiabatically expanding against pressure
from the IGM.  Detailed calculations with this model (Petitjean et al.
1993) show that parts of such an expanding cloud will have
temperatures that fall below $b=20$ \kms, though no part of such a
cloud will fall below 10 \kms.  Because the edges of the clouds are
cooler, there will be a correlation between column density and
temperature, which may explain the correlation seen between $b_{min}$
and N(H~I).  Different lines a sight through a particular cloud type
(key parameters are the central density of the cloud and it's
expansion velocity) trace out a curve in the N(H~I) - $b$ plane, with
a positive correlation between N(H~I) and $b$. The many different
cloud types required for consistency with global \Lya cloud properties
make it extremely difficult for this model to explain an overall
general correlation between N(H~I) and $b$ (such as was reported by
Pettini et al., but later shown to be an artifact of the line fitting
process by Rauch et al.) but a correlation between $b_{min}$ and
N(H~I) might be achievable, and would correspond to different lines of
sight through a cloud type with maximal central density and expansion
velocity.
 
Clouds may also cool by inverse Compton scattering of electrons on the
cosmic microwave background (Giallongo and Petitjean 1994), but their
calculations indicate that this will only cool \Lya forest clouds to
18.1 \kms.

Of course, the correlation between $b_{min}$ and N(H~I) may simply be
indicating that clouds with larger N(H~I) are more turbulent.
However, it is hard to see why this would lead to such a sharp lower
limit in the $b$ distribution, it does not explain how the lines below
N(H~I) $=10^{14}$ \cmm cool below 20 \kms, and it should lead to a
general correlation between N(H~I) and $b$, which we do not see.
A similar correlation between $b_{min}$ and N(H~I) has
been seen in hydrodynamical CDM simulations performed by Zhang,
Anninos, and Norman (1995).

\subsection {Possible Metal Line Contamination of \Lya Line List}

There are not enough unidentified metal lines in our line list to have
significantly affected any part of the $b$ distribution with measured
columns greater than $10^{12.5}$ \cmm.  In particular, there are many
more narrow lines in our line list than can be accounted for by
unidentified metals.

We can compare, per unit redshift, the number of metal lines inside
and outside the forest.  The lines outside of the forest are trivial
to identify -- any absorption seen is a metal line.  In general, a
metal line system will require more components to fit if it has been
observed with higher SNR.  Because the orders of our \object spectra
outside of the \Lya forest have much higher SNR than many of the
orders inside the \Lya forest, we can not simply compare the number of
metal lines inside and outside of the forest.  Instead we count groups
of lines.

We define a narrow line cluster as a group of one or more narrow lines
($b<20$ and $\rm{N(H I)} > 12.5$) grouped within 250 \kms, the scale
on which metal lines are known to cluster.  We found 74 clusters (254
narrow lines) in a wavelength interval of 1370 \AA \space above the
\Lya forest, all of which must be metals.  We would hence expect to
find 37 metal line clusters inside of the \Lya forest: we found 21 (65
lines).  We additionally found 35 clusters (47 lines) which we believe
to be \Lya.

This would indicate that half of the narrow lines we have identified
as \Lya may be metal lines.  However, we believe that most of
the 47 narrow lines identified as H~I \Lya are H~I for three reasons.
First, we expect that many of the narrow metal lines present inside of
the \Lya forest have been blanketed because many of the metal lines
outside of the forest are weak when compared with \Lya forest lines.
Second, the narrow lines we have identified as \Lya cluster much less
strongly on a 250 \kms scale than the lines we have identified as
metals (3.1 lines per cluster for identified metal lines vs 1.3 for
assumed \Lya).  Third, we did not find identifications for the narrow
\Lya cluster lines, even though we have coverage to C~IV and the
highest SNR to date.

\section {Implied He II optical depth of the \Lya forest}

Using the distribution functions we have found for the H~I \Lya
forest, we can calculate the expected He~II optical depth of the \Lya
forest as a function of N(He~II)/N(H~I).  The He~II optical depth at
$z=2.4$ and $z=3.3$ has been measured towards three high redshift QSOs
(Jakobsen et al. 1994, Davidson et al. 1996, Tytler \& Jakobsen
(unpublished)).  Since we wish to compare the expected He~II optical
depth of the \Lya forest with the observations, we will have to take
into account the redshift evolution of the \Lya forest.  We will
assume that the $b$ and N(H~I) distributions remain unchanged with
redshift, and that the normalization of the $z$ distribution evolves
like
\begin{equation}
{{dn} \over {dz}} \propto (1+z)^{2.6}.
\end{equation}
\noindent This is the redshift evolution found by Cristiani et
al. (1995), for H~I \Lya forest lines with N(H~I) $>10^{13}$ \cmm.
Because the N(H~I) distribution we found (at $<z> = 2.7$) agrees with
the N(H~I) distribution found by Lu et al. (1997) above N(H~I) $=
10^{12.3}$ \cmm at $<z> = 3.7$ (Lu et al. (1997) provided no
information about lower column lines), we know that lines with N(H~I)
$>10^{12.3}$ \cmm have the same redshift evolution.  There is no
information available on the evolution of lines with N(H~I)
$<10^{12.3}$ \cmm, so we will assume that they evolve like the lines
with N(H~I) $>10^{12.3}$ \cmm.  An example of simulated He~II \Lya
forest spectra at 7.9 \kms is shown in Figure 12.

To calculate the optical depth of the He~II \Lya forest, we have to
specify how much of the H~I $b$ value is thermal.  We have done the
calculation twice: once assuming that the H~I $b$ value is entirely
thermal, and once where 10 \kms of the H~I $b$ value is thermal with
the rest turbulent.  These two scenarios represent opposite extremes
in the possible $b$ values for He~II \Lya lines -- there is no known
way to cool an H~I \Lya forest line below 10 \kms.

The results of these calculations, with the N(H~I) distribution cut at
below $10^{12.1}$ \cmm, are presented in Figure 13.  In this case, the
\Lya forest can account for all of the observed He~II absorption at
$z=2.4$ when N(He~II)/N(H~I) $\approx 10^3$ to $10^4$, depending upon
how much of the H~I $b$ value is thermal.  These values are much
larger than expected from common models of the UV background radiation
(Miralda-Escude 1993b), but are not unreasonable if there is a large
break in the UV spectrum at the 228 \AA \space He II edge.

However, as pointed out by Songaila et al. (1995), if $\rm{N}_{min} <
10^{12.1}$ the \Lya forest can account for all of the observed He~II
absorption for much smaller values of N(He~II)/N(H~I).  In Figure 14,
we present the results of a calculation of \th with N$_{min} = 9.0$
(effectively zero).  In this case, N(He~II)/N(H~I) needs only be
$\approx 100$ to $\approx 150$ for the \Lya forest to account for all
of the observed He~II optical depth at $z=2.4$.  The difference
between thermal and turbulent H~I $b$ values is now less significant
because unsaturated lines are now responsible for a much larger
fraction of the \th.

We would like to determine how much of the high redshift He~II optical
depth is due to the \Lya forest, because we would then know how much
is due to the IGM (insomuch as any optical depth not due to the forest
must come from the IGM, i.e. the Gunn-Peterson effect).
Unfortunately, after viewing both Figure 13 and Figure 14, it becomes
clear that it is entirely possible for the \Lya forest to be
responsible for all of the observed high redshift He~II optical depth,
or a small fraction of it (or anywhere in between).  To
ultimately determine the He~II optical depth of the \Lya forest, we
will have to determine N(He~II)/N(H~I) in the forest (probably via
observations the He~II lines of \Lya forest clouds).  We will
also have to determine N$_{min}$ or place a limit on it that is less
than $10^{9}$ \cmm, as well as the redshift evolution of the low
N(H~I) lines.  Determining N$_{min}$ will be extremely difficult.

In any event, unless N(He~II)/N(H~I) is much lower than expected, \th
will be dominated by either the IGM or low N(H~I) \Lya forest clouds.
The redshift evolution of \th will hence give the redshift evolution
of either low N(H~I) \Lya forest clouds, the IGM, or a combination of
both.

Since the number of observed lines per \AA~ of spectrum increases like
$(1+z)^{2.6}$ (Equation 5), and the equivalent width of each line goes
as $(1+z)$, we expect that the redshift evolution of the He~II optical
depth from \Lya forest clouds to be given by
\begin{equation}
\tau(z)_{\rm{HeII}} \propto (1+z)^{3.6}.
\end{equation} 
\noindent 
The calculations in Figures 12 and 13 indicate that this is indeed the
case and that the blanketing effects of saturated He~II lines do not
strongly effect the redshift evolution of the \Lya forest \th.

The key point is that the redshift evolution of the \Lya forest He~II
optical depth is dominated by the number of lines present at any
redshift.  It does not depend upon whether or not the H~I $b$ values
are thermal.  It does not depend upon N(He~II)/N(H~I) (it does depend
upon any change of N(He~II)/N(H~I) with redshift).  It does not
depend upon the form of the N(H~I) distribution below $10^{13}$ \cmm.
Hence if \th is found not to have the redshift evolution found in
Equation 6, either the IGM or the low N(H~I) \Lya forest lines are
evolving differently than the high N(H~I) \Lya forest lines.

Based on the measurement of $\tau_{\rm{HeII}} = 1.0$ at $z=2.4$ by
Davidsen et al., if the He~II optical depth is due entirely to the
\Lya forest \th should be 2.4 at $z=3.3$.  If this value of \th is
found, it would imply that the IGM and the \Lya forest are composed of
the same type of objects, a view supported by numerical cosmological
simulations (Hernquist et al. 1996).  If this value of \th were found
it would still be possible for there to be a physically
distinguishable IGM, but it's redshift evolution would have to be the
same as the redshift evolution of the \Lya forest.  A widespread
ionization of the diffuse gas would affect both the IGM and the forest
in the same way, and could produce the sort of redshift evolution seen
in the forest.  Alternatively, a value of \th $\approx 1.7$ at $z=3.3$
(the lower limit found by Jakobsen et al. 1994) would imply that
either the low column \Lya forest and/or the IGM are evolving less
rapidly than the high column \Lya forest, or that the ionizing
spectrum is harder at high $z$, reducing N(He~II)/N(H~I).  A harder
spectrum at high redshift is not expected because the UV opacity
increases with redshift.  Songaila \& Cowie (1996) noticed that 
SiIV/C IV decreases rapidly at redshifts above 3.1, which they 
account for as due to a softening of the ionizing spectrum due to the 
IGM becoming optically thin to He I ionizing photons; additional data
from Boksenberg (private communication), however, does not support this trend.

\section {Line Clustering}

To investigate any possible clustering of the forest lines we calculated
the two point correlation
\begin{equation}
{\xi(v)} = {{{N(v)} \over {N_{exp}(v)}} - 1}, 
\end{equation}
\noindent where $N(v)$ is the number of line pairs with velocity
separation $v$ in the line list, and $N_{exp}(v)$ is the number of
line pairs expected if the lines were randomly placed along the line
of sight.  $N_{exp}(v)$ and the 1 $\sigma$ error bars were calculated
via Monte Carlo simulations. For each echelle order in the real
spectrum, the number of lines found in that order were randomly placed
in the redshift range covered by the order.  $N_{exp}(v)$ was then
calculated from the line list containing the randomly placed lines
from all orders.  The process was repeated 2,500 times to get an
average value for $N_{exp}(v)$ and a reliable standard deviation.  We
randomly placed lines within each echelle order as opposed to the
entire spectrum because we are primarily interested in clustering on
small velocity scales, and we wanted to avoid any effects that may be
caused by the fact that the higher SNR orders generally have more
lines than the lower SNR orders.  One order covers a velocity range of
$\approx 4000$ \kms.

The $\xi(v)$ function for all 466 \Lya forest lines in our spectrum is
presented in Figure 15. The drop below 50 \kms indicates that our
spectrum is missing 106 line pairs below 50 \kms.  Since the easiest
ways to lose a line pair at low velocity separations is to have one
line blanket another or to have two lines blend into one line, we
believe that we are missing approximately 106 lines from our line list
due to blending and blanketing (assuming that there is no intrinsic
clustering).  The missing lines will have parameters distributed like
the observed distributions, so the most likely N(H~I) of a missing
line is about $10^{12.5}$ \cmm (Figure 3), and the most likely $b$
value is about 23 \kms (Figure 4).  The lines that were blanketed will
have been dropped from the line list, but some lines that are missing
because of blending will have significantly altered the $b$ value of
the line that was observed.  There should be negligible blending of
lines that are separated by more than 50 \kms because the $\xi(v)$
does not show any deficit of lines at those separations.  Because the
components of blends had N(H~I) $\approx 10^{12.5}$ and a $b \approx
23$ \kms, the blended lines should have N(H~I) $>10^{12.5}$ and $b$
values of at most 50 + 23 \kms $\approx 70$ \kms.  The 106 missing
lines are hence capable of producing much of the ``hump'' in the
observed N(H~I)-$b$ distribution (Figure 5).  However, because it
would be difficult for these blended lines to produce a line with
$b>70$ \kms, it seems unlikely that the tail of the observed $b$
distribution above 70 \kms can be accounted for by blending.  This
lends credence to our suspicions that the extended non-Gaussian tail
above $b = 80$ \kms in the observed distribution is real or a result
of continuum placement errors in the real data.

We see no evidence for significant clustering on any scale.  Over the
velocity range $50 < v < 150$ \kms we find $<\xi> = 0.06 \pm 0.045$,
which is 5.8 standard deviations from the value of $0.32 \pm 0.08$
found by Webb (1987), and 2.4 standard deviations from the value of
$0.17 \pm 0.045$ found by Hu et al. (1995) over the same velocity
range.

The lack of clustering in the entire line sample might be because our
line list contains considerably more lines at lower column densities
than the other studies, and it is suspected that lines with higher
N(H~I) cluster more strongly (Crotts 1989; Chernomordik 1995;
Cristiani et al. 1995).  However, $\xi(v)$ for the 84 \Lya lines with
$10^{13.6} < \rm{N(H~I)} < 10^{14.5}$ (presented in Figure 16) still
does not show significant clustering on any scale $> 50$ \kms.  Our
value of $<\xi>$ = $0.19 \pm 0.24$ over $50 < v < 150$ \kms is
significantly less (2.3 $\sigma$) than the value of $0.73 \pm 0.13$
found by Hu et al. for lines in the same N(H~I) range.  These lines do
cluster in the 25-50 \kms bin, and our value of $<\xi>$ over $25 < v <
150$ \kms (same range as before, but now including the 25-50 \kms bin)
is $0.57 \pm 0.21$.  Neither Rauch et al. (1992) or Lu et al. (1996)
found clustering on any scale in the \Lya forest.  We do not
understand why some \Lya forest studies found strong clustering while
others do not.

\section {The Dispersion of the $b$ distribution}

\Lya lines have different intrinsic $b$ values because their
temperatures differ or their turbulent velocities differ.  Clustering
on small velocity scales will also change $b$ values.  If all of the
H~I $b$ values measured in \Lya forest clouds are thermal, simple
photoionization models in which the clouds are in thermal equilibrium
and heated only by the UV background flux predict that the baryonic
content of the \Lya forest would far exceed the total baryonic content
of the universe (Press and Rybiki 1993), because the lines with the
largest $b$ values would be very highly ionized and hence must be very
massive to account for their observed H~I.  Hence, the gas is
turbulent. If the gas is heated by a means other than the UV
background, the H~I/H~II ratio will generally be lower than for gas
heated by photoionization, exacerbating the problem.  Spectra of the
metal lines associated with moderate column (N(H~I) $\approx 10^{13}$
to $10^{14}$ \cmm) \Lya clouds (SNR $\approx 100$) may be able to
measure the turbulent velocities of the clouds, but the issue is
currently unresolved.

Several authors (Hu et al. 1995, de la Fuenta et al. 1996) have
suggested that most of the \Lya forest is at the same temperature, and
the entire dispersion observed in the $b$ distribution is due to
strong clustering on velocity scales of 10 \kms.  We call this
scenario the isothermal hypothesis.  If it is correct, then most of
the lines in our line list are blends, and higher SNR observations
will resolve some of the blends.  This will have several results on
the observed distributions.  The $\xi(v)$ function will increase on
small velocity scales -- it may not rise above unity, but it will
increase relative to it's present value.  The observed $b$
distribution will become narrower, and the peak of the $b$
distribution may move to a smaller $b$.  With this in mind, some
credence may be lent to the isothermal hypothesis by the fact that the
peak of our $b$ distribution is 5 \kms less than that found by Hu et
al. (1995), a study similar to ours but with 0.5 to 0.75 the SNR of
our data.  On the other hand, we find a greater dispersion in the
observed $b$ distribution than Hu et al. (1995), which is inconsistent
with the clustering hypothesis.

The lack of observed clustering below 50 \kms in the full line list is
not necessarily inconsistent with the hypothesis. Blending of close
lines could wipe out any clustering signature, such that the only
clustering signature would be the shape of the $b$ distribution.
Because it is unclear exactly how the fitting process will treat a
clustered intrinsic distribution, simulated-observations will be
required to determine the effect of small scale clustering on the
observed two point correlation function.  Because of the time
involved, we have not performed any simulated-observations in which
the \Lya clouds are clustered on small velocity scales.  

\section {Redshift Evolution of the Intrinsic Distributions}

There are two other studies that have determined the intrinsic
distributions of the \Lya forest using the same type of analysis that
we have used.  Hu et al. (1995) found the intrinsic distributions at
$<z> = 2.8$, and Lu et al. (1996) determined the intrinsic
distributions at $<z> = 3.7$.  The SNR of both studies were less than
that of our observations, 50 for Hu et al. (1995) and 20-80 for Lu
et al. (1996).

We have found (at $<z>=2.7$) the N(H~I) and $b$ distributions, as well
as the clustering properties of the clouds, to be nearly identical to
that found by Lu et al. (1996) at $<z> = 3.7$.  However, as previously
noted, we found a different $b$ distribution and different clustering
properties than Hu et al. (1995), which was done at the same redshift
as our study.  Although we do not understand why Hu et al. (1995)
found different cloud properties than we did, we suspect that the
differences in the derived intrinsic distributions stem from different
procedures used to fit the absorption lines, because we both used
simulated-observations to remove SNR effects from the data.

Both Lu et al. (1996) and ourselves used the VPFIT software to fit
absorption lines, and we both used simulated-observations to determine
the intrinsic distributions of the \Lya forest.  Because of this we
believe that a meaningful comparison can be made between our results,
and we conclude that there is almost no redshift evolution of the \Lya
forest (other than in the number of absorbers).  The only apparent
difference is in the dispersion of the $b$ distribution, which we
found to be 14 \kms whereas Lu et al. (1996) found it to be 8 \kms.
Although we have not done the simulations to measure the error on our
value of 14 \kms, but we believe the error may be as large as 3 or 4
\kms -- and that the difference between our value and that reported by
Lu et al. (1996) is not necessarily significant.

\acknowledgments

We thank the referee, L. Lu, for many helpful comments.  We thank
J. Aycock and J. Killian at the W.M. Keck Observatory for help during
the observations.  D. Welty found metal identifications for several of
the narrow lines in the spectra.
We thank T. Barlow for providing a copy of his code
that automatically extracts and wavelength calibrates HIRES data.  We
thank R. Carswell for providing a copy of his Voigt profile fitter,
VPFIT.  We thank S. Rappoport for aid with the simulations.  We thank
P. Petitjean, and M. Rauch and especially S. Burles for very helpful
discussions.  This work was supported in part by grants AST-9420443
and NAGW-4497.

\clearpage

\figcaption { The spectrum of QSO \object.  We indicate 466 \Lya
	forest lines and 73 metal lines.  The spectrum is normalized
	to a continuum level of one.  Wavelengths are vacuum
	heliocentric values.  The log column density (in log
	cm$^{-2}$) and velocity dispersion (in \kms) of each line are
	displayed above a tick mark indicating the center of the line.
	Identified metal ions are indicated along with the redshift of
	the metal system.  The calculated 1$\sigma$ error spectrum
	(per pixel) is shown just above zero on the same scale.  The
	pixels size is 2 \kms.}

\begin{figure}
\figurenum{1}
\plotone{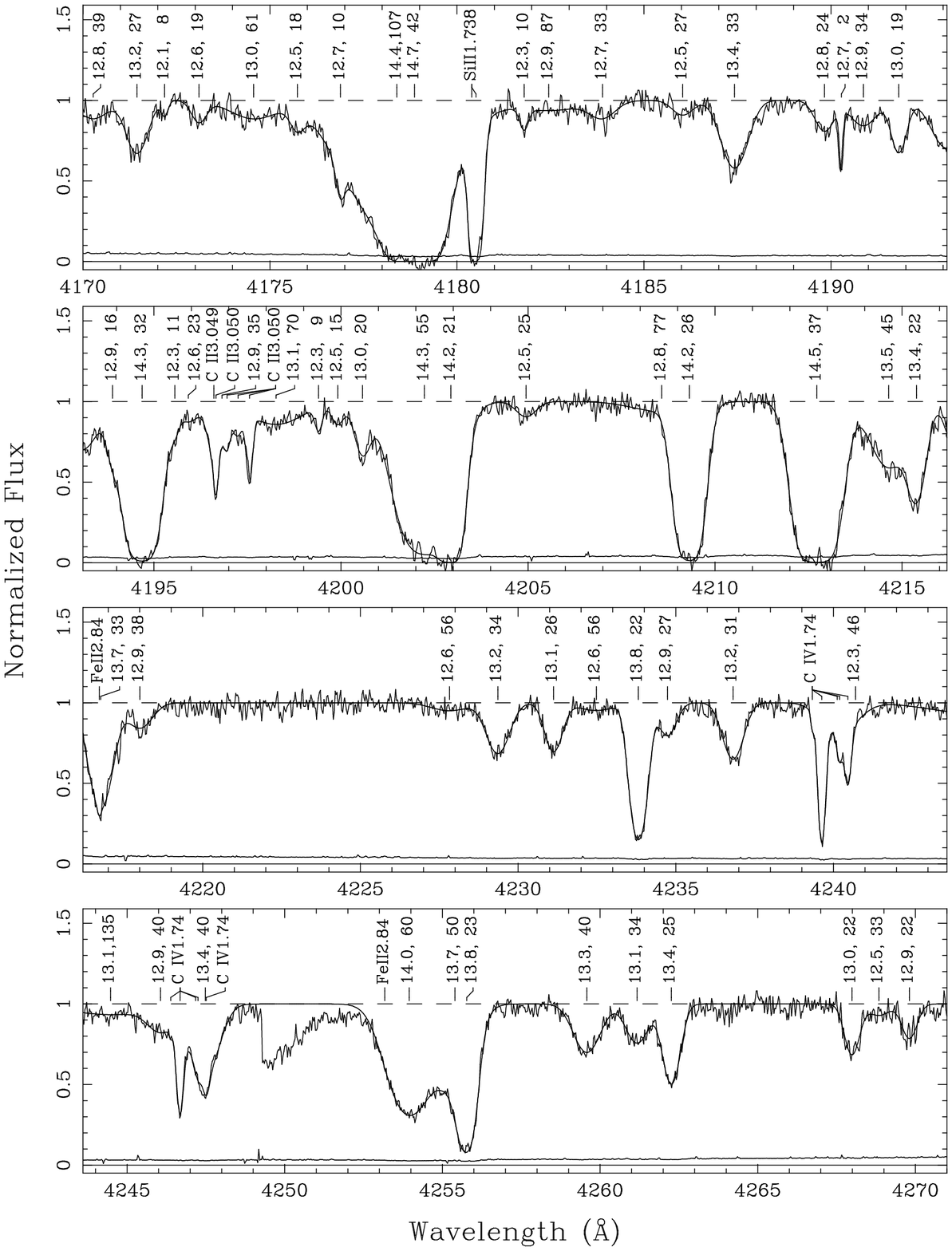}
\end{figure}

\begin{figure}
\figurenum{1}
\plotone{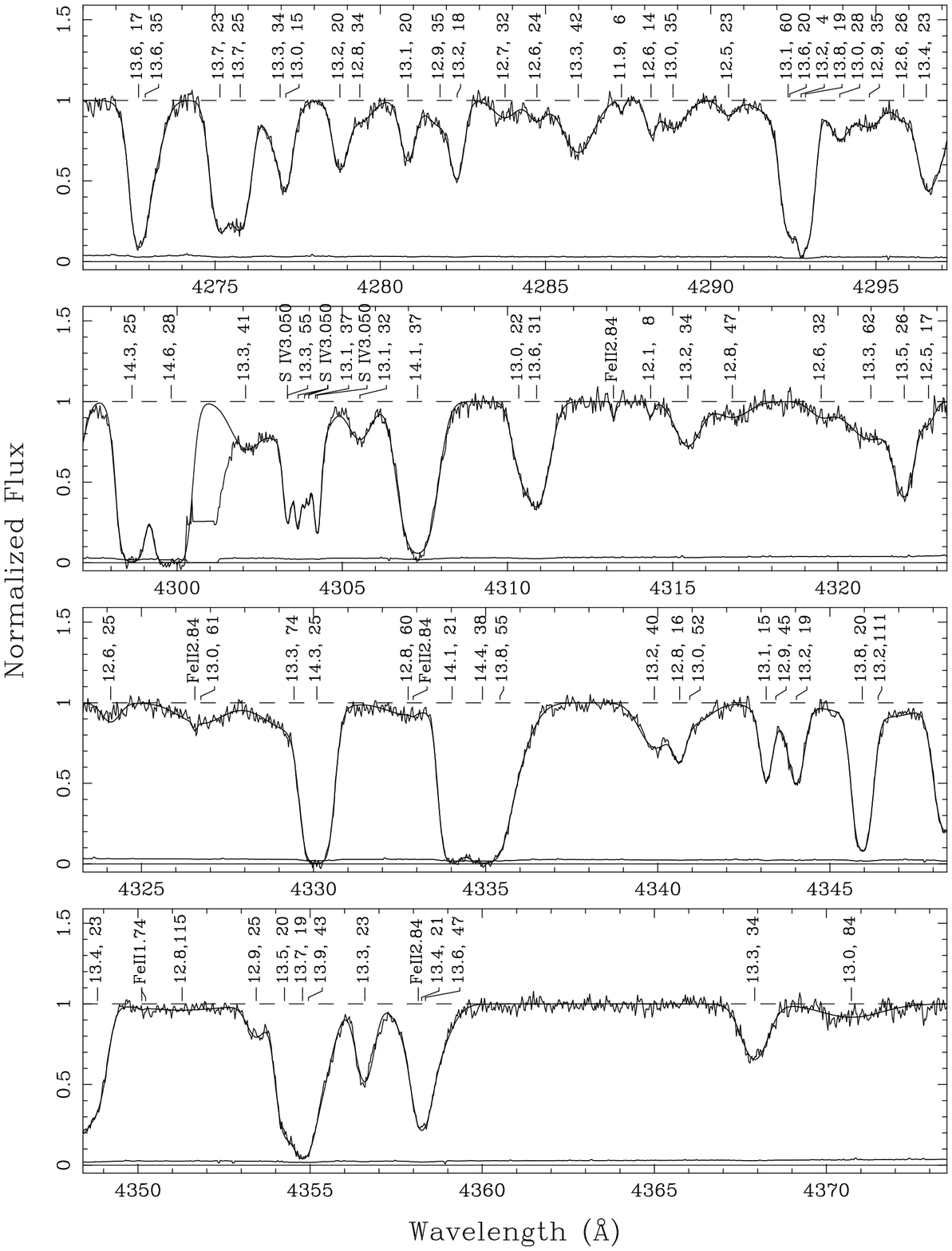}
\end{figure}

\begin{figure}
\figurenum{1}
\plotone{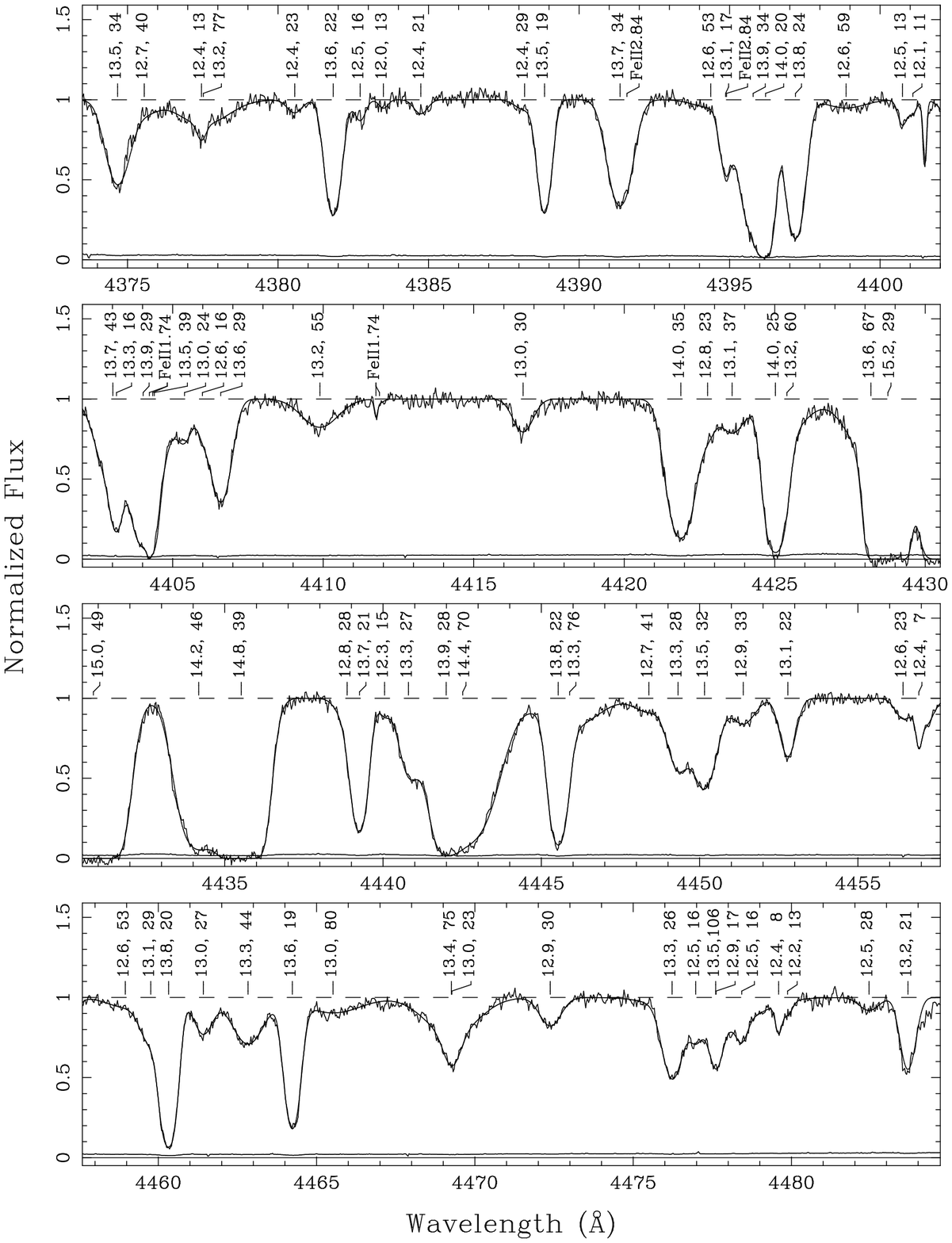}
\end{figure}

\begin{figure}
\figurenum{1}
\plotone{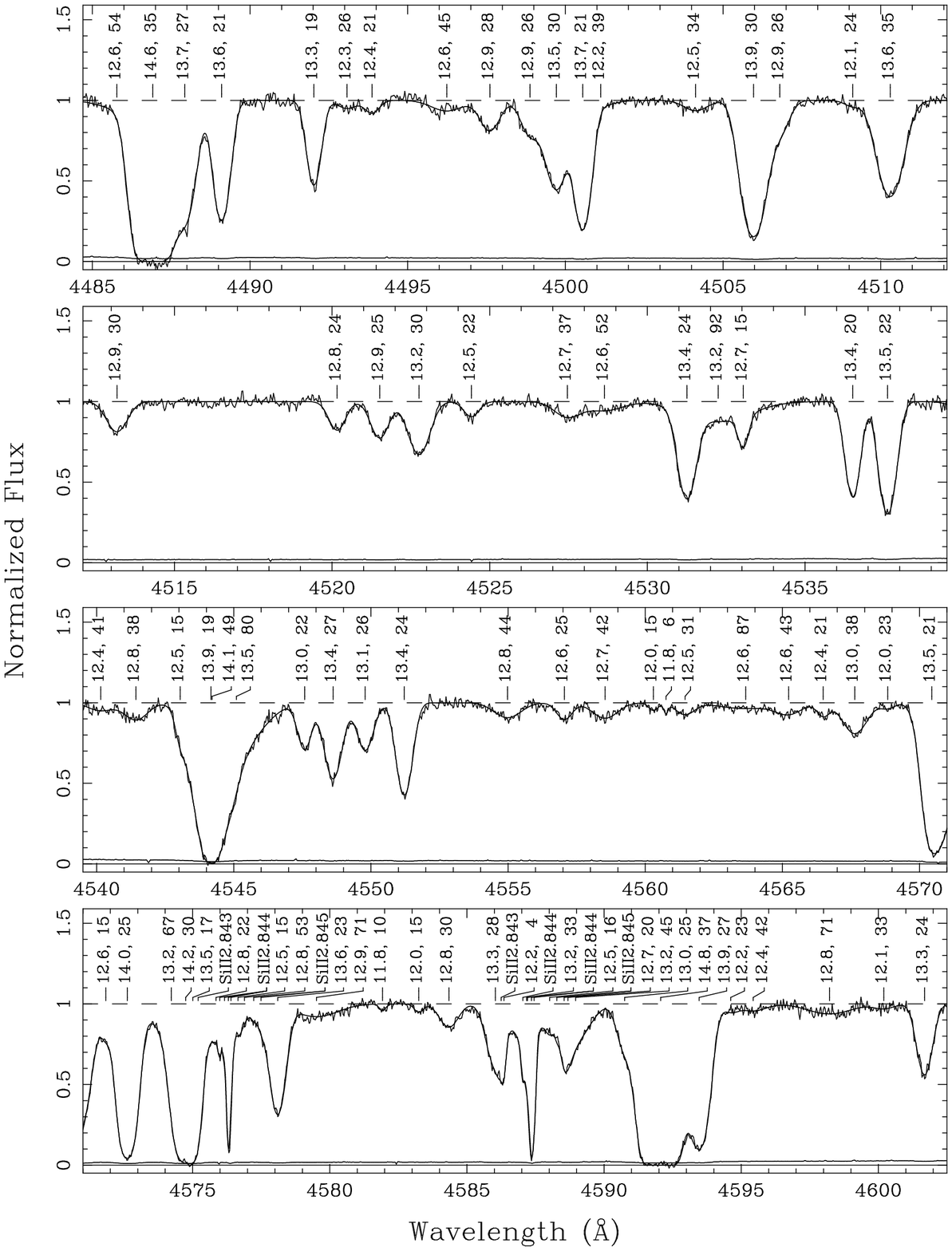}
\end{figure}

\begin{figure}
\figurenum{1}
\plotone{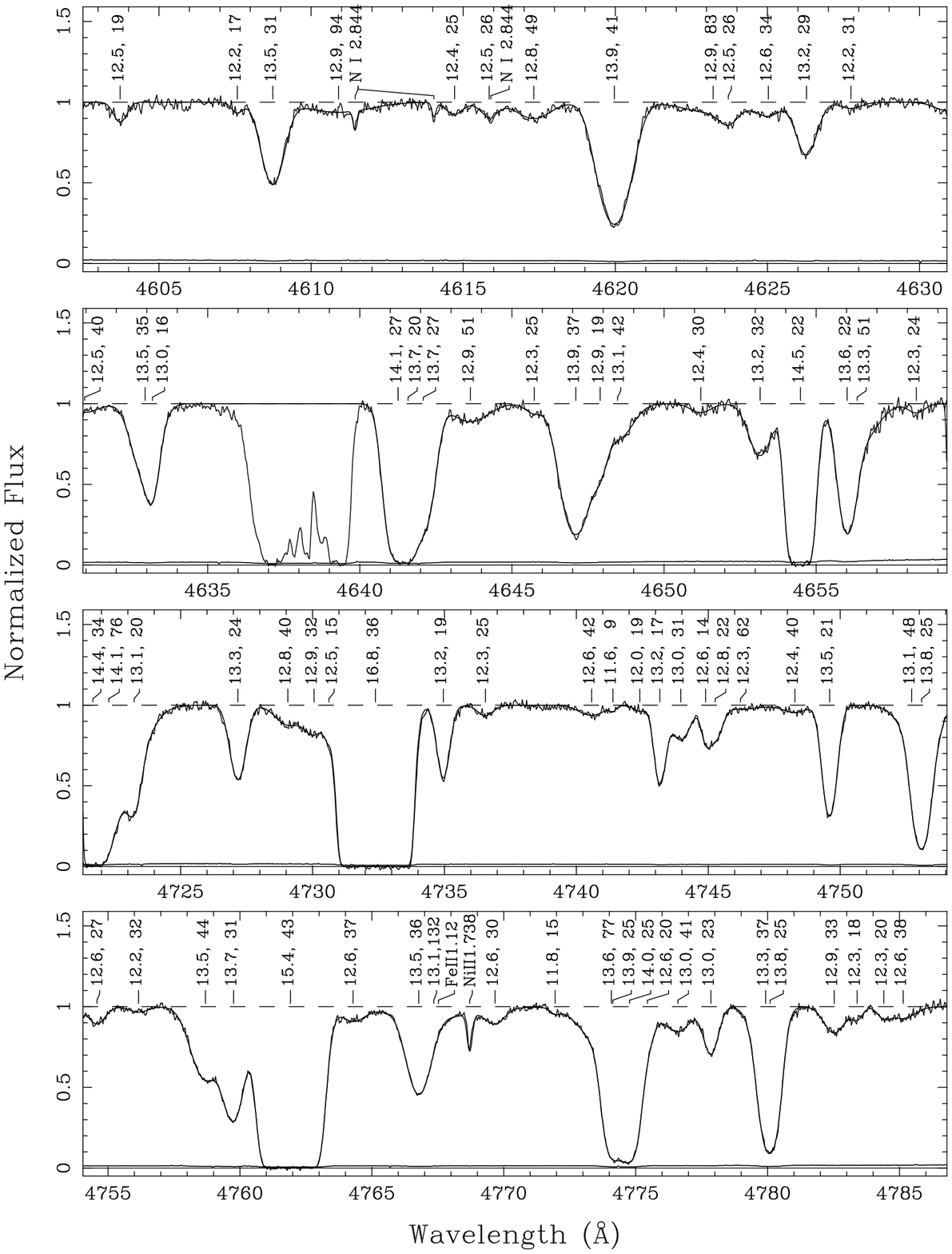}
\end{figure}

\begin{figure}
\figurenum{1}
\plotone{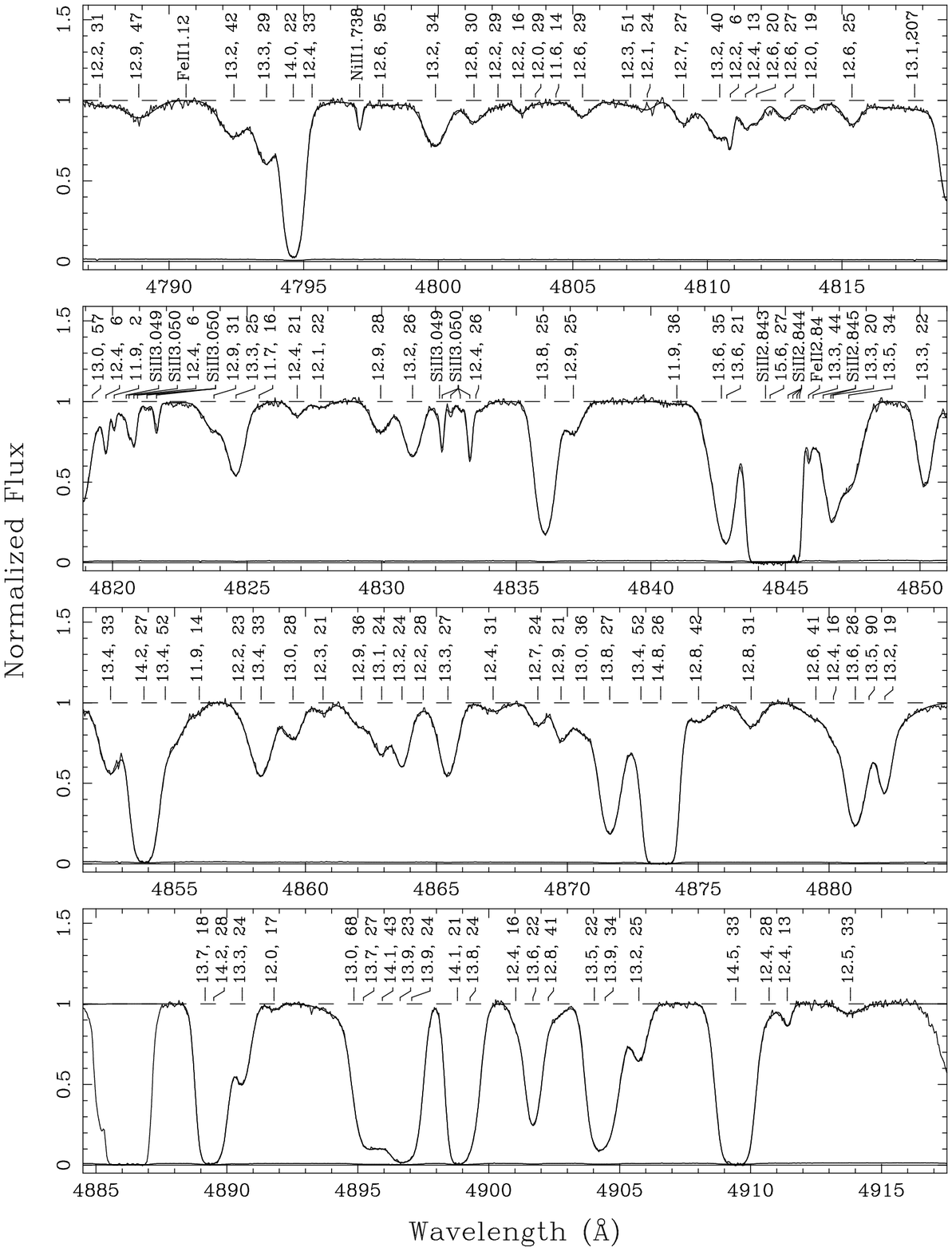}
\end{figure}

\begin{figure}
\figurenum{2} 
\plotfiddle{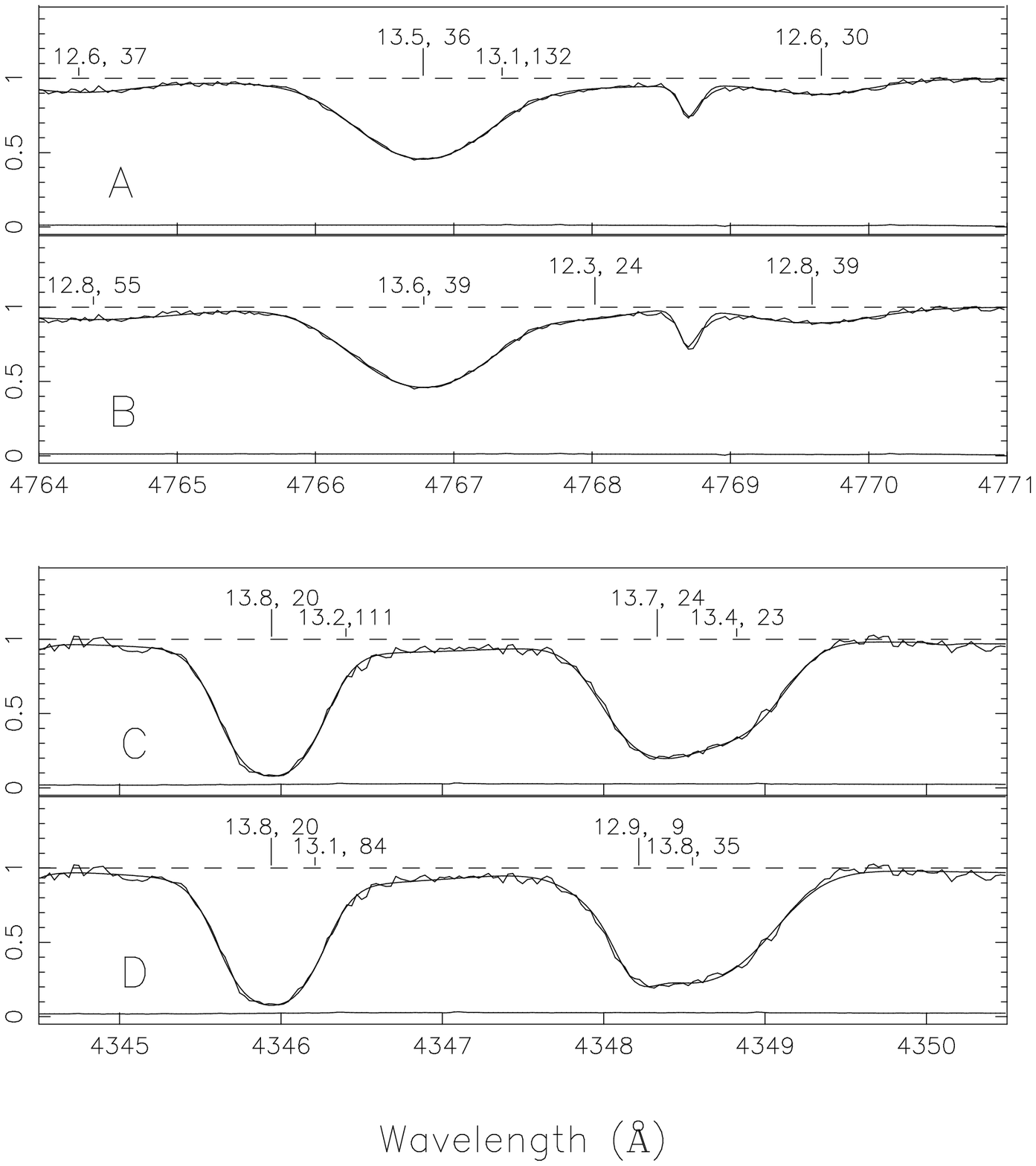}{5.0in}{0.0}{70.0}{70.0}{-225.0}{-50.0}
\caption { Examples of the non-uniqueness of the line fitting
	process.  Panels A and B show the same region of spectra fit
	by different sets of Voigt profiles; panels C and D show a
	different region of spectra.  The paired fits were started
	with slightly different parameter, which led to a different
	local minima.  The line near 4768.7 \AA \space is an Fe II
	(z=1.12) line constrained by other transitions, and was not
	varied during the fitting process.  The reduced $\chi^2$ for
	each region are as follows: Panel A -- 0.83, B -- 1.15, C --
	1.04, D -- 1.16.}
\end{figure}

\begin{figure}
\figurenum{3}
\plotfiddle{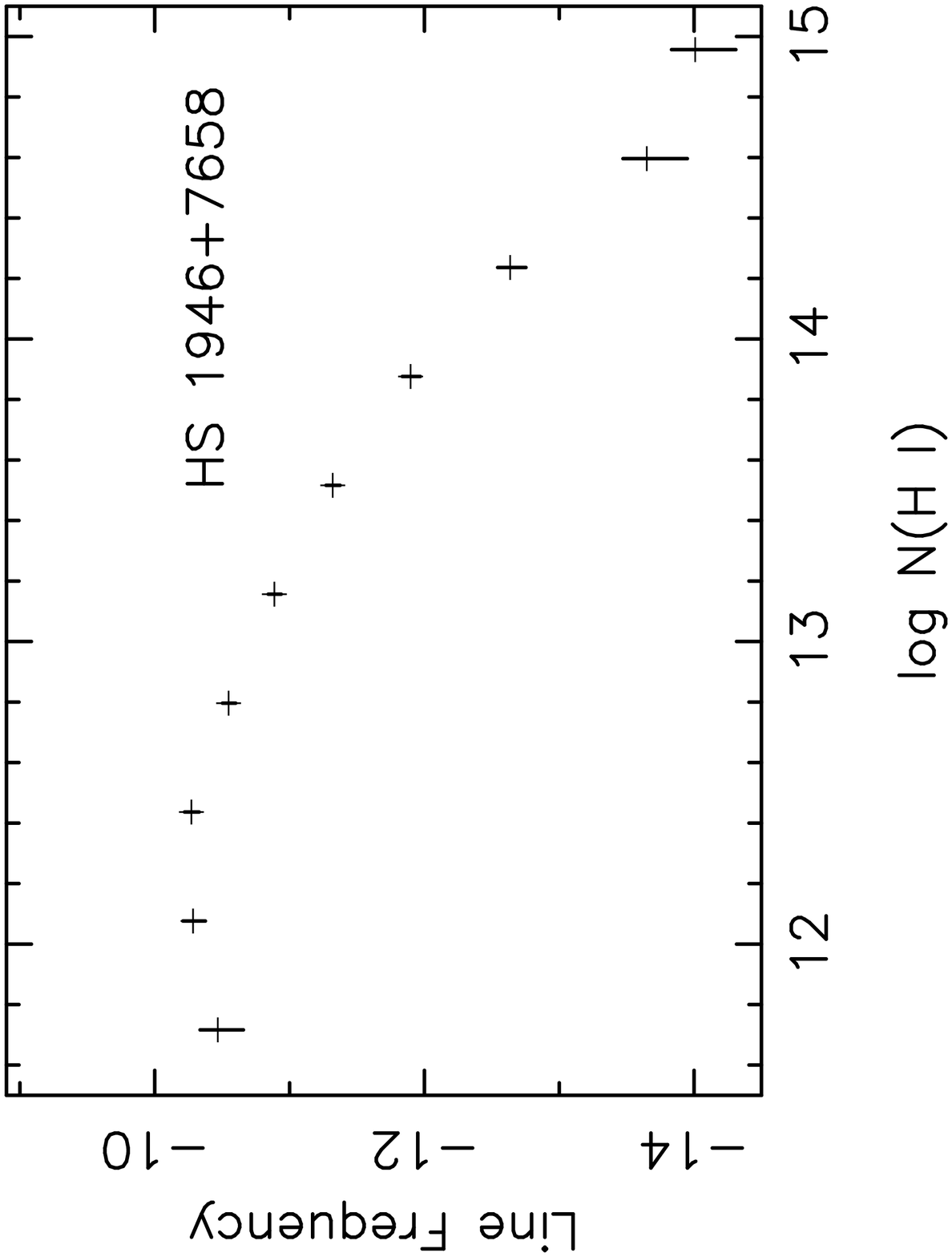}{5.0in}{-90.0}{70.0}{70.0}{-280.0}{400.0}
\caption { The observed N(H~I) distribution towards HS 1946+7658.
	The lines have been binned into intervals of 0.36 in Log
	N(H~I), starting from 11.75. The vertical bars represent 1
	$\sigma$ errors.  The first bin has 8 lines.}
\end{figure}

\begin{figure}
\figurenum{4}
\plotfiddle{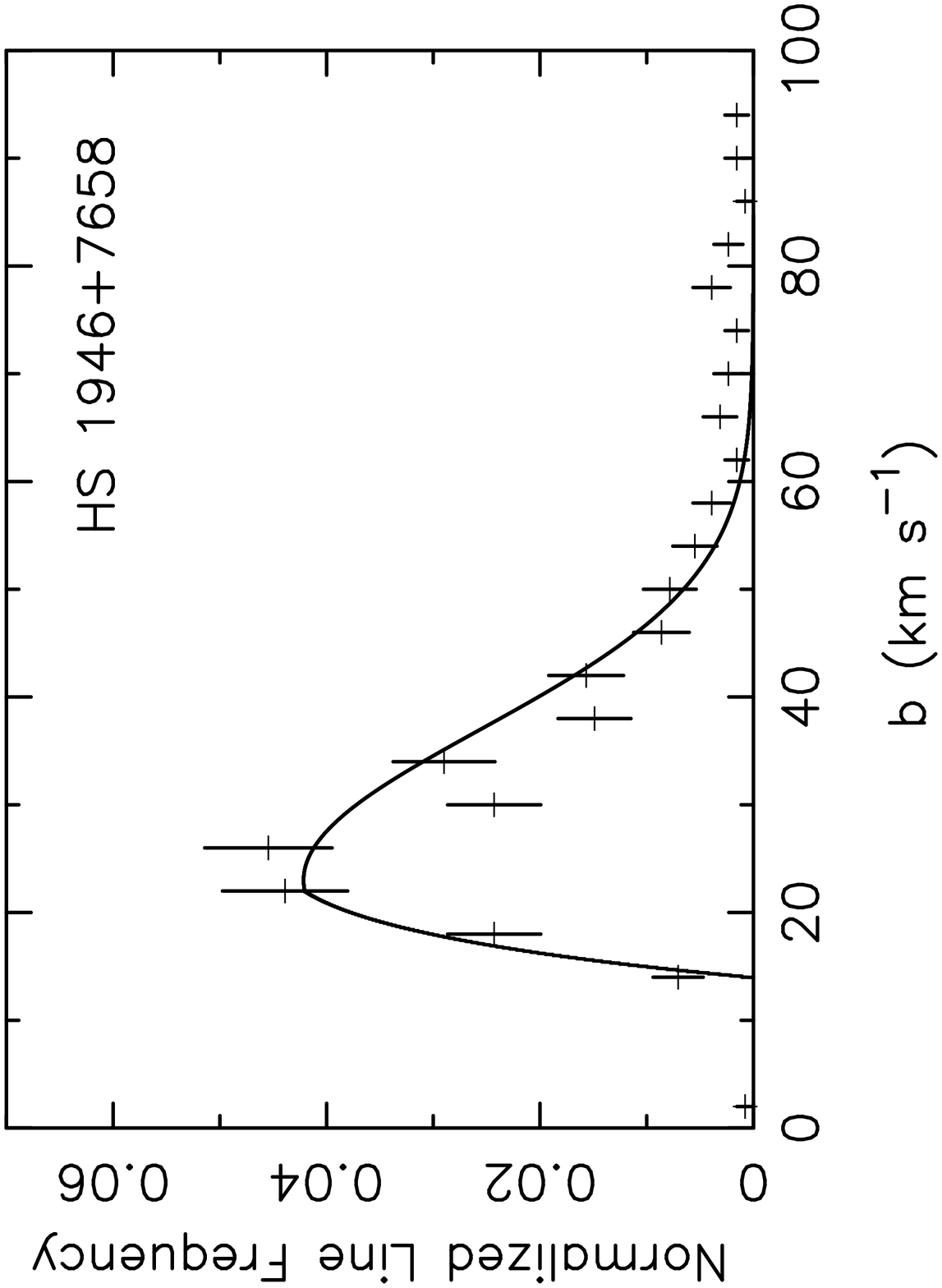}{5.0in}{-90.0}{70.0}{70.0}{-280.0}{400.0}
\caption { The observed $b$ distribution towards HS 1946+7658.  The
	lines have been binned into intervals of 4 \kms in $b$.  The
	vertical bars represent 1 $\sigma$ errors.  The curve is our
	best estimate of the intrinsic $b$ distribution -- a Gaussian
	with mean 23 \kms, $\sigma_b = 14$ \kms, and $b_{min}$ given
	by Equation 4. We believe that most of the lines with $b<14$
	\kms are not part of the intrinsic $b$ distribution.  The
	lines with $b>60$ \kms are probably real, but some must be
	blends and some may be continuum errors. The sum of all line
	frequencies is one.}
\end{figure}

\begin{figure}
\figurenum{5}
\plotfiddle{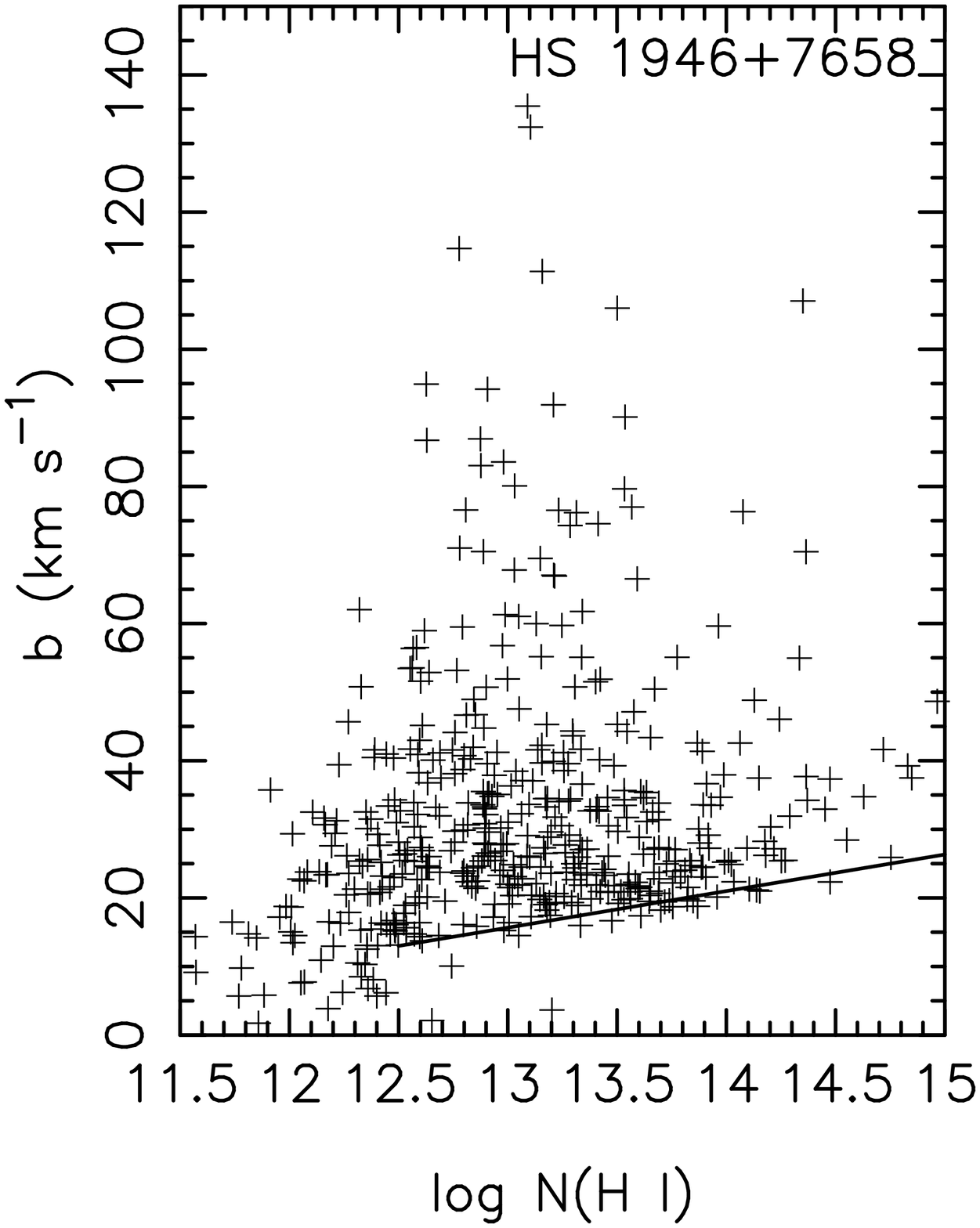}{5.0in}{0.0}{70.0}{70.0}{-225.0}{-50.0}
\caption { The observed N(H~I)-$b$ distribution of \Lya lines
	towards HS 1946+7658.  The position of each line is indicated
	by a cross.  Errors have not been displayed. They are listed
	in Table 1.  The solid line is the lower cutoff in the $b$
	distribution described in Section 3.6 (Equation 4).  The 3
	lines below this cuttoff with N(H~I) $> 10^{12.5}$ are all
	real.  They may be metals (Section 3.7).}
\end{figure}

\begin{figure}
\figurenum{6}
\plotfiddle{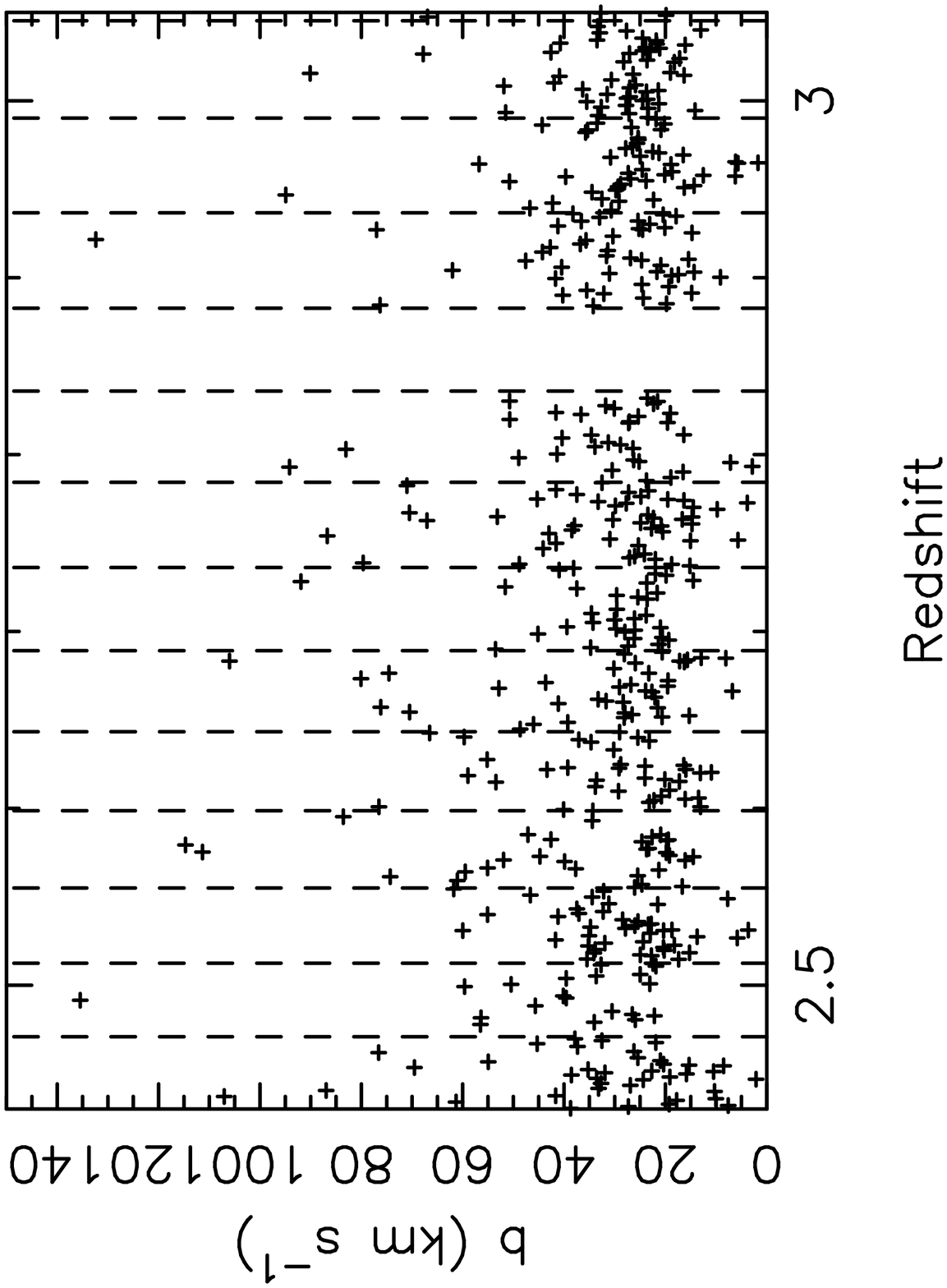}{5.0in}{-90.0}{70.0}{70.0}{-280.0}{400.0}
\caption { The observed $z$-$b$ distribution towards HS 1946+7658.
	The boundaries of the echelle orders are indicated by dashed
	vertical lines.}
\end{figure}

\begin{figure}
\figurenum{7}
\plotfiddle{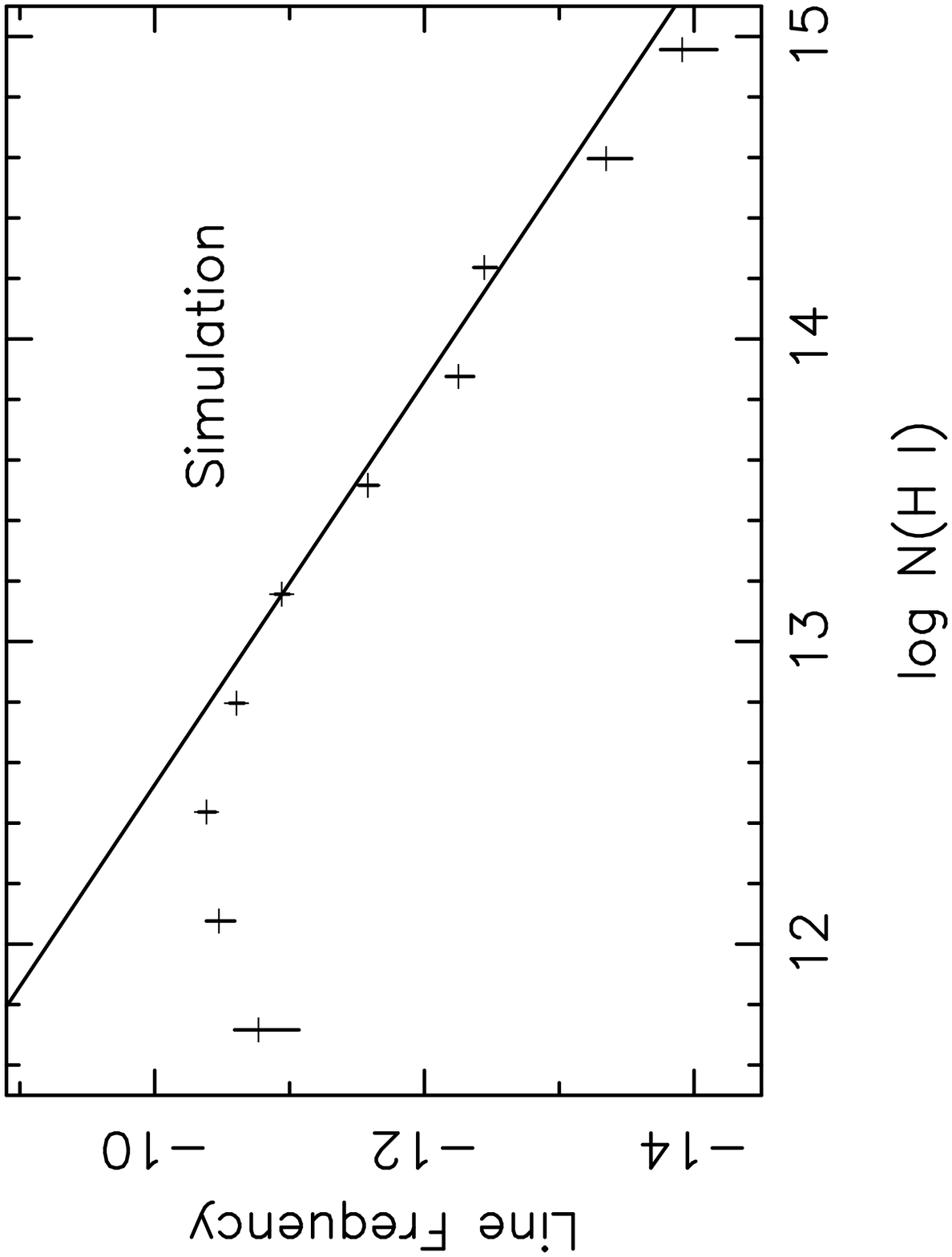}{5.0in}{-90.0}{70.0}{70.0}{-280.0}{400.0}
\caption { Simulated-observations. The intrinsic N(H~I)
	distribution (straight line, given by Equation 2) and the
	results of the simulated-observations, binned into intervals
	of 0.25 in Log N(H~I).  The vertical bars are 1 $\sigma$
	errors in the simulated-observations.  }
\end{figure}

\begin{figure}
\figurenum{8}
\plotfiddle{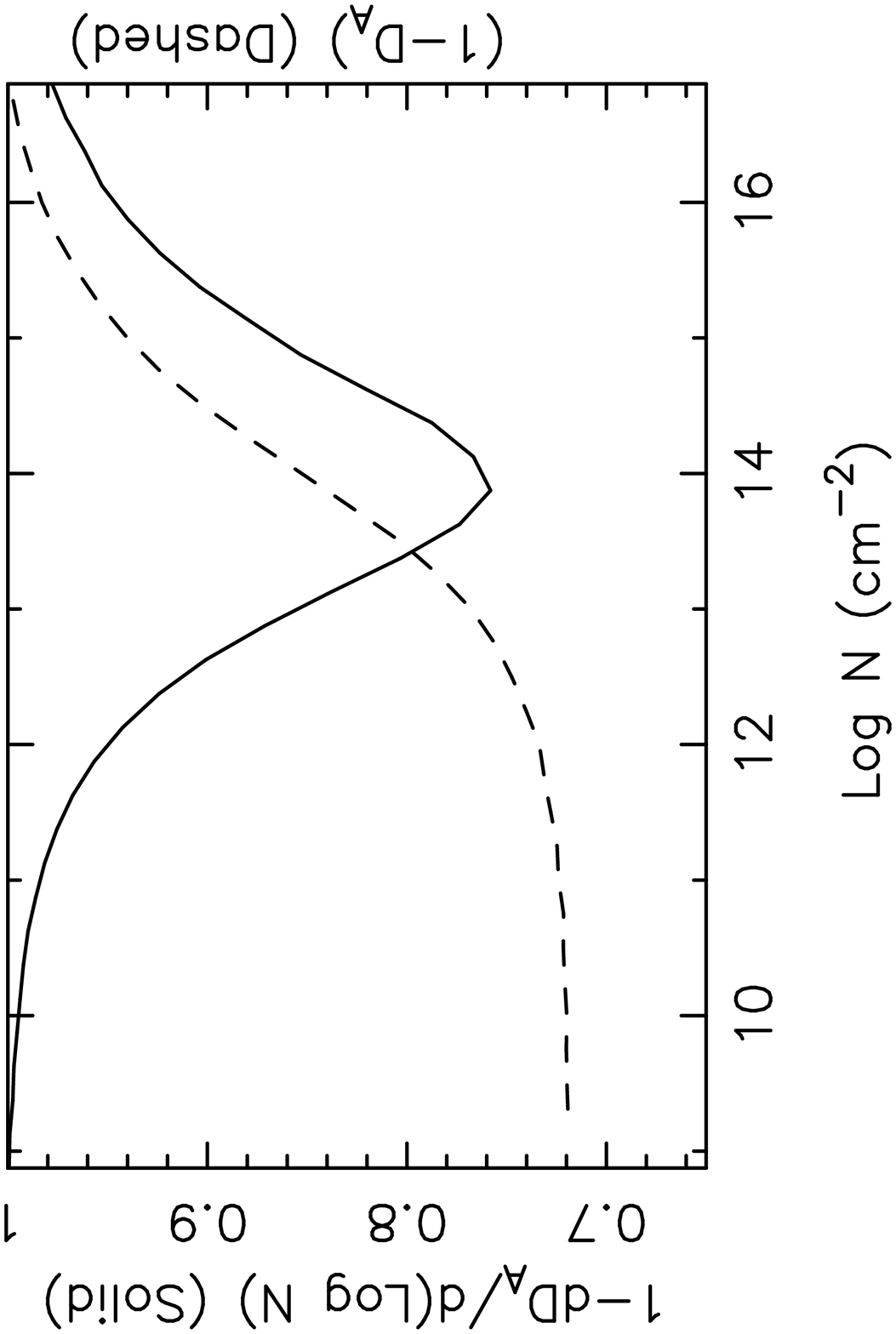}{5.0in}{-90.0}{70.0}{70.0}{-280.0}{400.0}
\caption { The dashed line shows the mean H~I \Lya absorption
        (D$_A$) from an intrinsic population with a N(H~I)
        distribution given by Equation 2 at N(H~I) $<10^{17}$ \cmm
        plotted against N$_{min}$ of the distribution.  This ignores
        continuum absorption.  The solid line shows the differential
        mean H~I \Lya absorption of the same population of clouds as a
        function of N(H~I).  Note that most of the H~I \Lya forest
        opacity occurs in clouds with $10^{13} < \rm{N(H~I)} <
        10^{15}$., and that the differential contribution to the \Lya
        opacity is a maximum at N(H~I) $=10^{14}$ \cmm.}
\end{figure}

\begin{figure}
\figurenum{9}
\plotfiddle{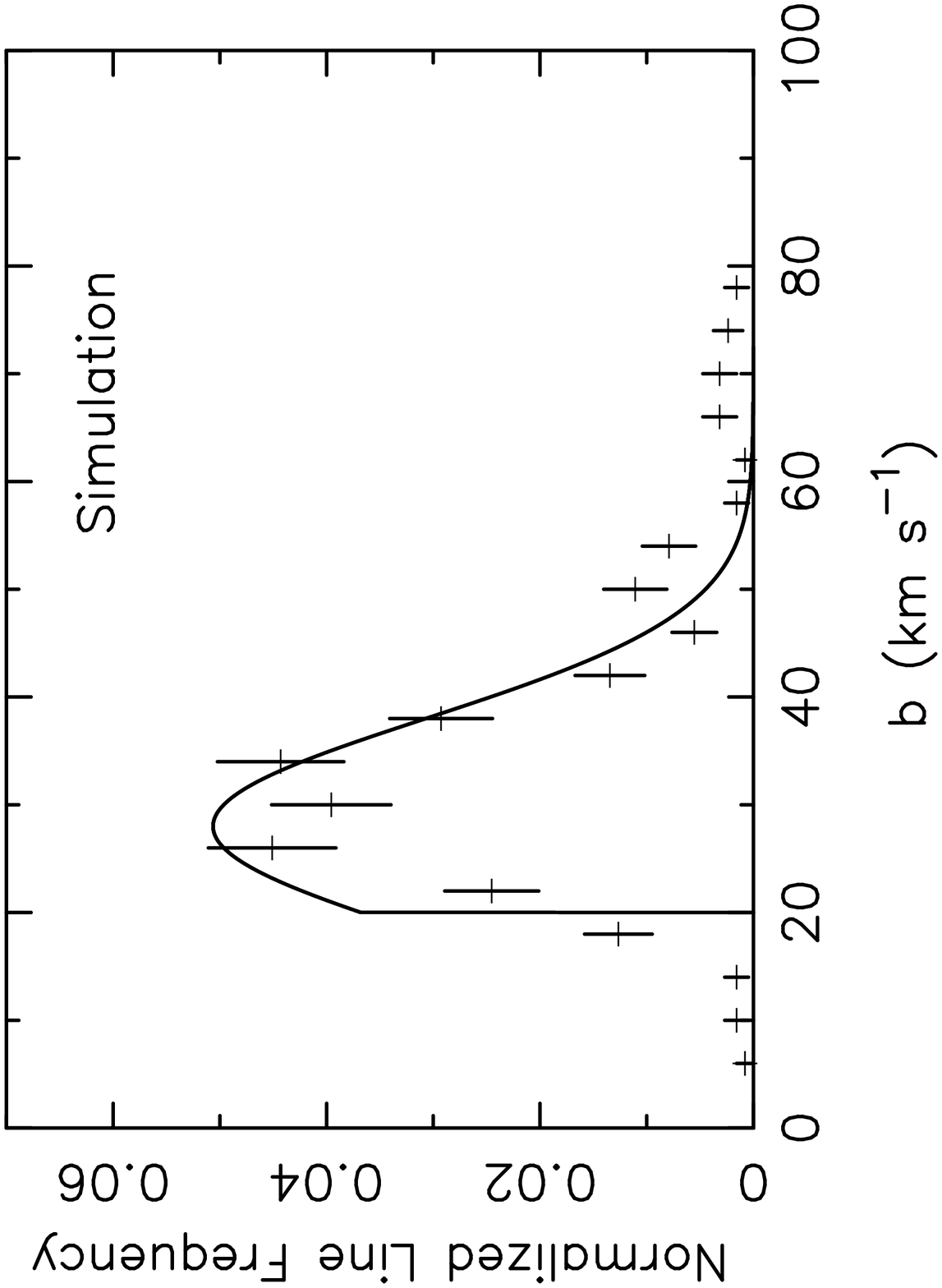}{5.0in}{-90.0}{70.0}{70.0}{-280.0}{400.0}
\caption { The simulated-observation $b$ distribution and the
	intrinsic $b$ distribution for the simulation.  The lines were
	binned into intervals of 4 \kms in $b$.  The vertical bars are
	1 $\sigma$ errors in the simulated-observations.  The
	intrinsic distribution for this simulation (not our best
	estimate of the true intrinsic distribution) is a truncated
	Gaussian with mean 28 \kms and $\sigma_b = 10$ \kms (displayed
	as a solid curve).  Note that the distribution of ouput $b$
	values are very similar to the input.}
\end{figure}

\begin{figure}
\figurenum{10}
\plotfiddle{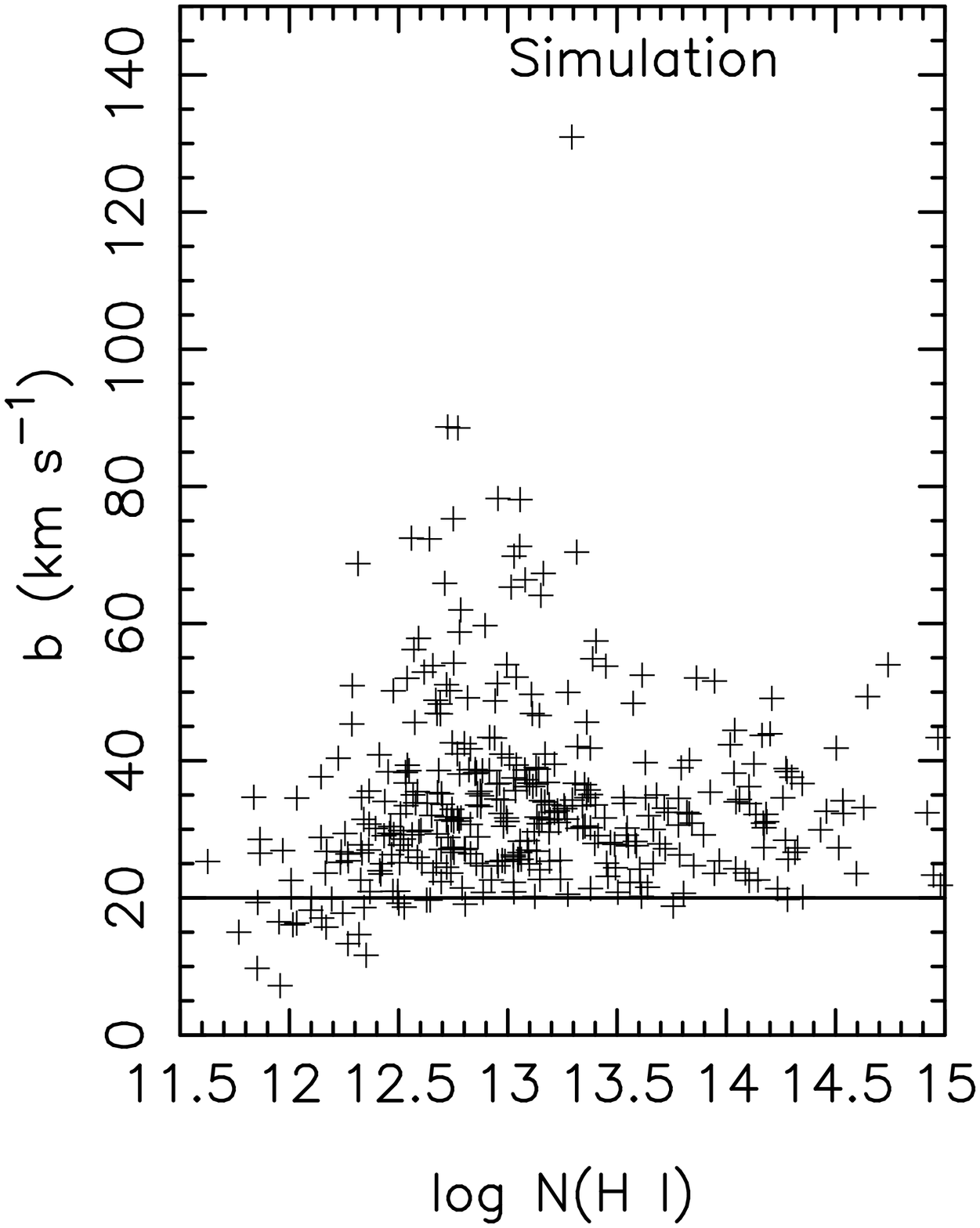}{5.0in}{0.0}{70.0}{70.0}{-225.0}{-50.0}
\caption { The simulated-observation N(H~I)-$b$ distribution.
	The lines of the simulated-observation are indicated
	by crosses.  The horizontal line at 20 \kms was the lower cutoff
	in the intrinsic distribution used to created the simulated 
	observation.  Note that this lower cutoff is completely preserved for
	lines with N(H~I) $>10^{12.5}$ \cmm, but that noise features
	with $b<20$ \kms appear as spurious lines at N(H~I) $< 10^{12.5}$.}
\end{figure}

\begin{figure}
\figurenum{11}
\plotfiddle{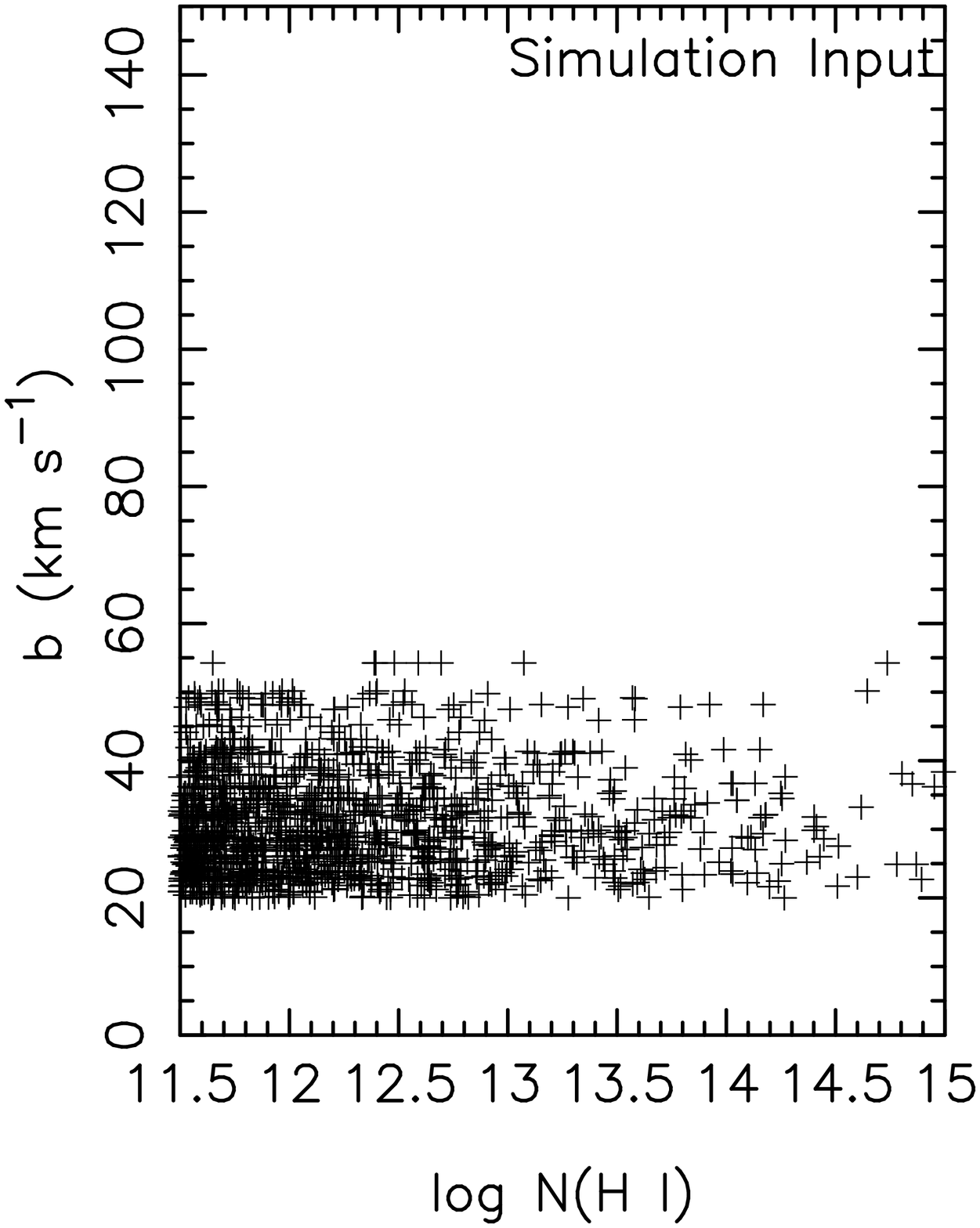}{5.0in}{0.0}{70.0}{70.0}{-225.0}{-50.0}
\caption { The intrinsic N(H~I)-$b$ distribution used to create the 
	data set the simulated-observations were made from.  Note the
	sharp lower cuttoff of 20 \kms in the $b$ distribution.}
\end{figure}

\clearpage
\begin{figure}
\figurenum{12}
\plotone{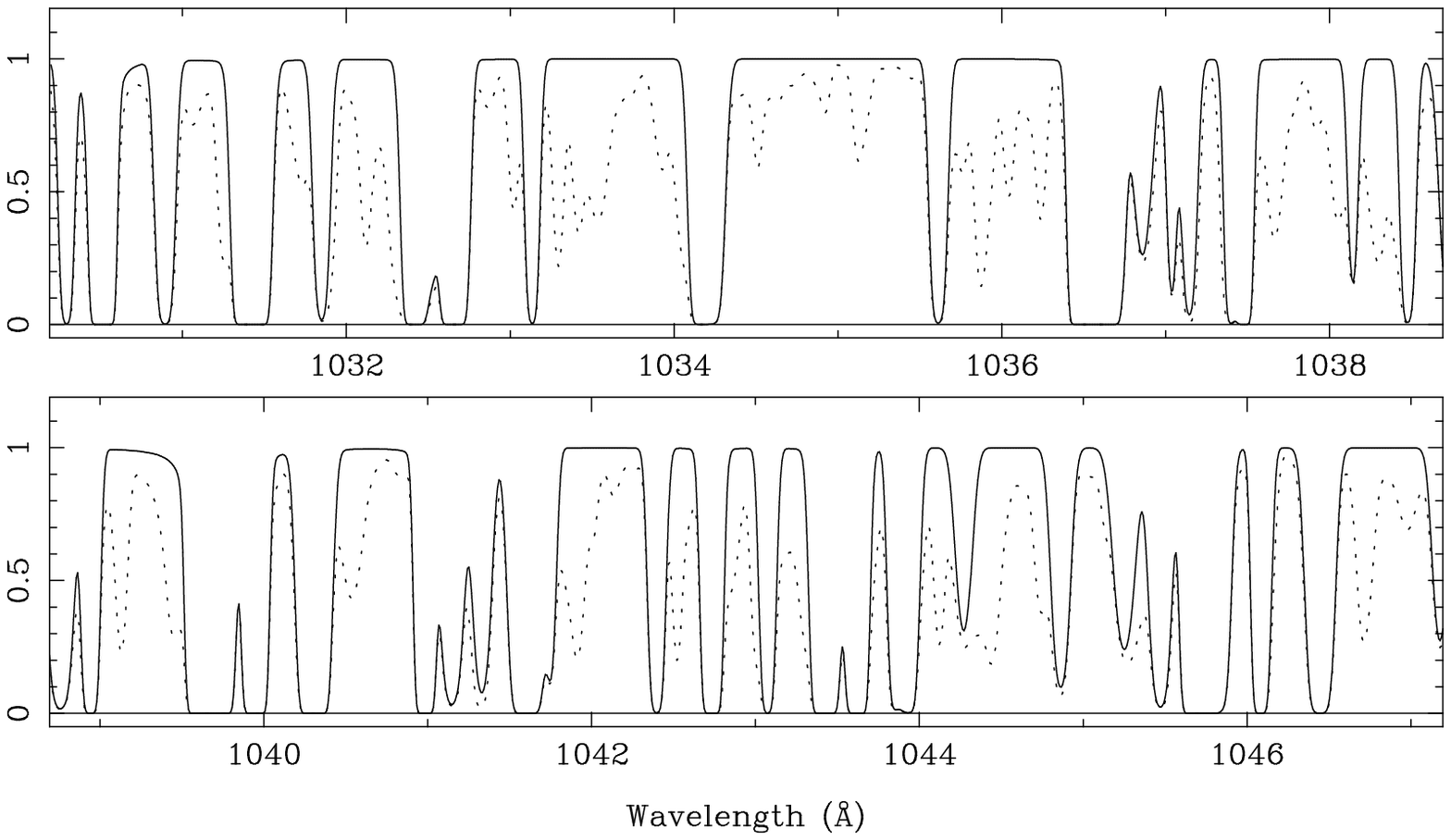}
\caption { A section of simulated 7.9 \kms He~II \Lya forest
	spectra at $z=2.4$.  This spectra was created using the H~I
	\Lya forest line distributions, with N(He~II)/N(H~I) = 100.
	The H~I $b$ values were taken to be thermal.  The solid line
	shows the spectra of \Lya forest clouds with N(H~I) $>
	10^{12.1}$, the dotted line shows the spectra of all lines
	with N(H~I) $> 10^{9.0}$ (effectively zero).  No noise has
	been added to the spectrum.  Note that even the spectra with
	all lines with N(H~I) $>{9.0}$ contains regions with very little
	absorption.}
\end{figure}

\begin{figure}
\figurenum{13}
\plotfiddle{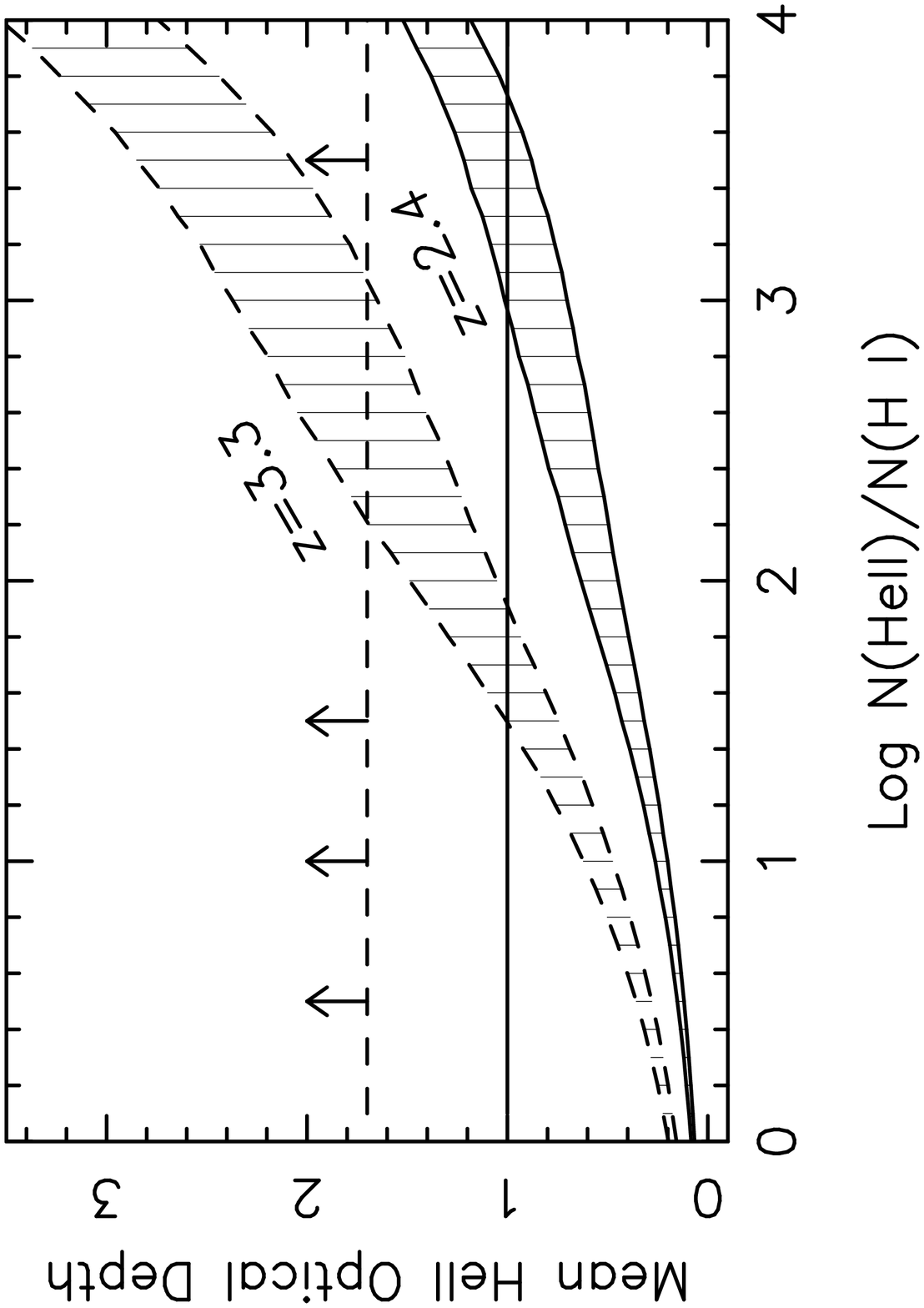}{5.0in}{-90.0}{70.0}{70.0}{-280.0}{400.0}
\caption { Mean He~II \Lya forest optical depth with $\rm{N}_{min}
	= 10^{12.1}$ at $z=3.3$ (dashed) and $z=2.4$ (solid). The
	lower curve for each redshift was calculated assuming the H~I
	$b$ values are entirely thermal, and the upper curve was
	calculated assuming mostly turbulent $b$ values (see Section 4
	for details).  The horizontal lines indicate observed
	He~II optical depths.  (1.0 at $z=2.4$, and a lower limit of
	1.7 at $z=3.3$) Note that if the He~II optical depth is due
	entirely to the \Lya forest, \th must be 2.4 at $z=3.3$.
	if N(He~II)/N(H~I) is constant with $z$.}
\end{figure}

\begin{figure}
\figurenum{14}
\plotfiddle{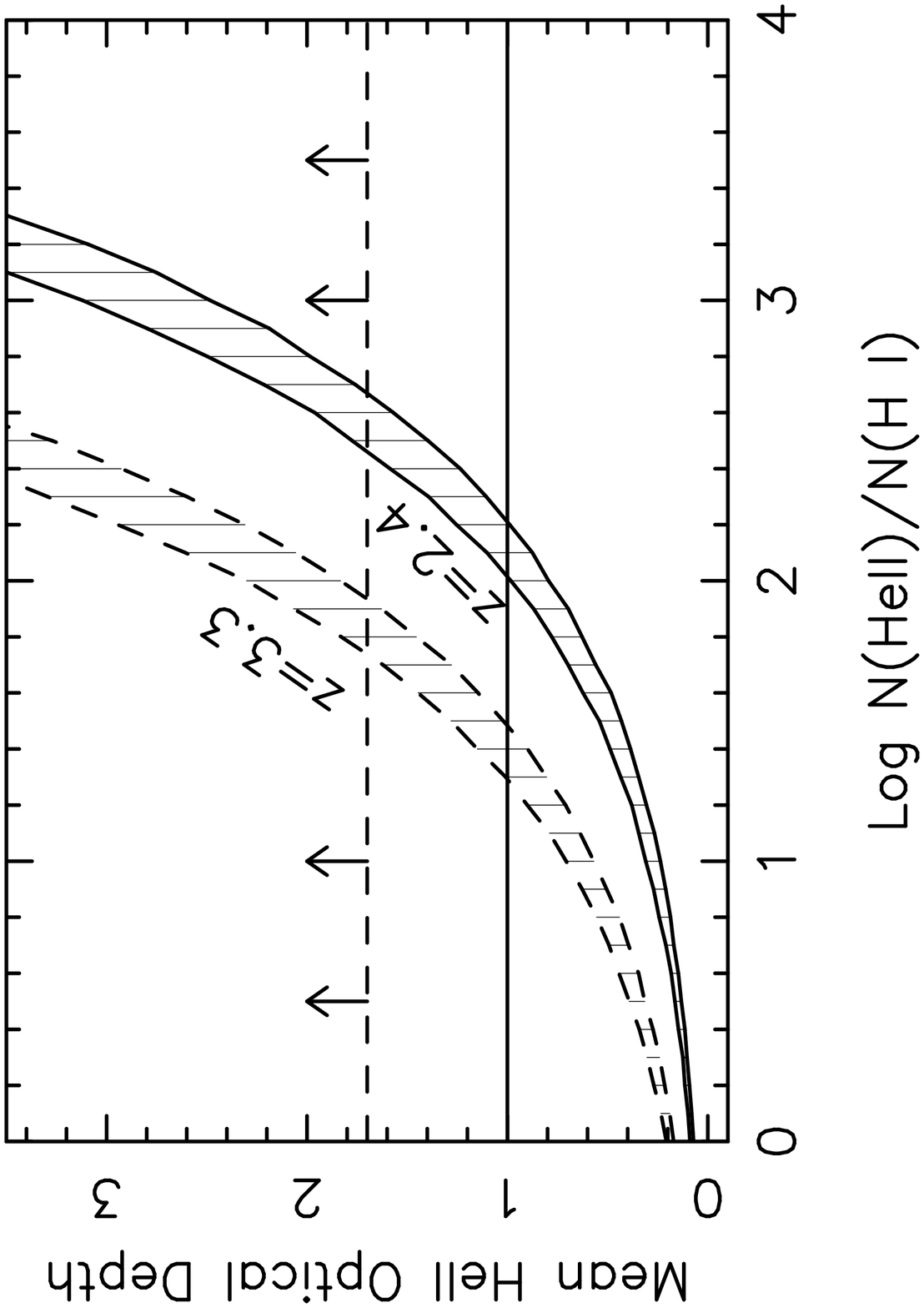}{5.0in}{-90.0}{70.0}{70.0}{-280.0}{400.0}
\caption {  Same as Figure 11 but with $\rm{N}_{min} = 10^{9.0}$ \cmm.}
\end{figure}

\clearpage
\begin{figure}
\figurenum{15}
\plotfiddle{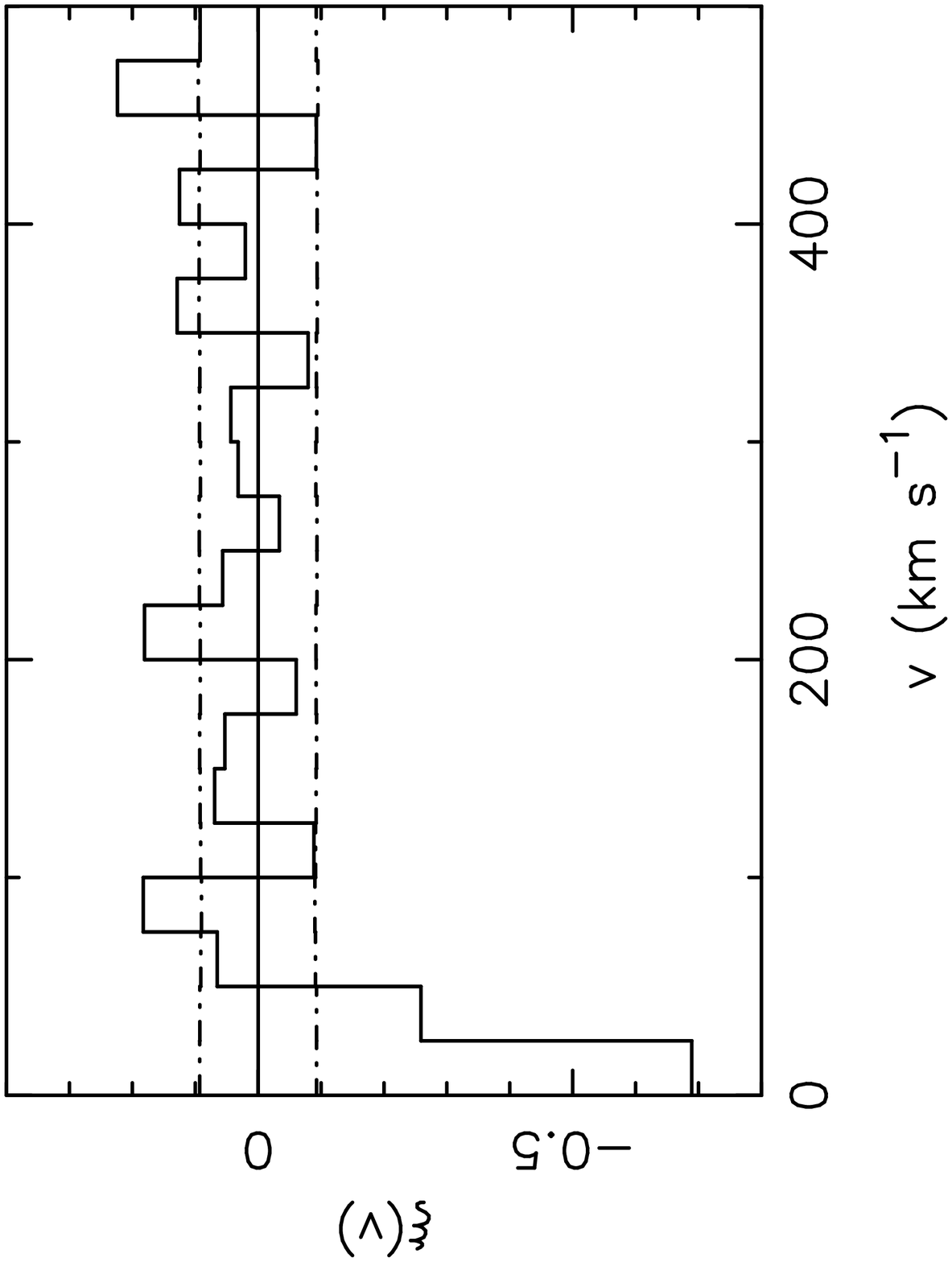}{5.0in}{-90.0}{70.0}{70.0}{-280.0}{400.0}
\caption { The two point correlation function for the 466 \Lya
	forest lines in the spectrum of HS 1946+7658.  The bin size is
	25 \kms.  The dashed line is the 1 $\sigma$ error array.  The
	decrease at $v < 50$ \kms is due to line blending and
	blanketing.}
\end{figure}


\begin{figure}
\figurenum{16}
\plotfiddle{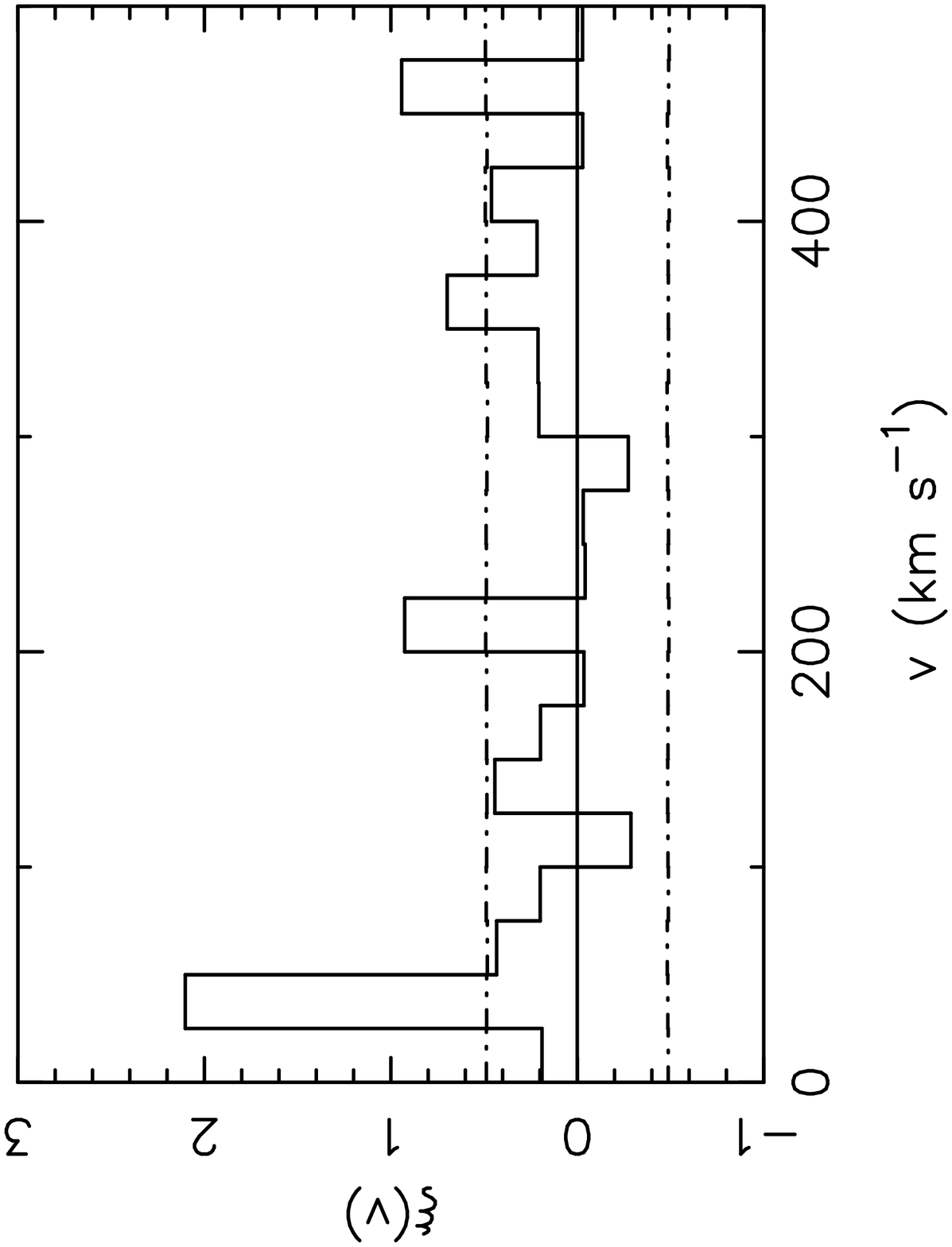}{5.0in}{-90.0}{70.0}{70.0}{-280.0}{400.0}
\caption { The two point correlation function for only the 120 \Lya
	forest lines in the spectrum of HS 1946+7658 with N(H~I)
	$>10^{13.5}$.  The bin size is 25 \kms.  The dashed line is
	the 1 $\sigma$ error array.}
\end{figure}

\end{document}